\documentclass[11pt]{article}

\usepackage[T1]{fontenc}
\usepackage[utf8]{inputenc}
\usepackage{textcomp}
\usepackage{amsmath}
\usepackage{amssymb}
\usepackage{graphicx}
\usepackage{url}
\usepackage[margin=1in]{geometry}
\usepackage{setspace}
\usepackage{longtable}
\usepackage{booktabs}
\usepackage{caption}
\usepackage[section]{placeins}
\usepackage{xcolor}
\usepackage[super]{natbib}
\bibpunct{}{}{,}{s}{}{}
\title{\textbf{Active-Space Quantum Simulation of N$_2$ Hydrogenation at a Ru Single-Atom Site on Ru(0001)}}

\author{Geet Gupta \\  \normalsize\itshape QpiAI India Private Ltd., Yashoda Nagar, Jakkur, Bengaluru \\ \\\normalsize Email: \textcolor{red}{geet.g@qpiai.tech}}

\date{}

\let\oldcite\cite
\renewcommand{\cite}[1]{\textcolor{green}{\oldcite{#1}}}

\begin{document}

\maketitle

\begin{abstract}
\noindent Selective activation of dinitrogen (N$_2$) under mild conditions is difficult: N$\equiv$N is one of the strongest bonds in chemistry, and most heterogeneous catalysts capable of breaking it require high temperature and pressure. Atomically dispersed transition-metal sites offer a computationally tractable route to studying the strongly correlated intermediates involved. We connect periodic DFT calculations with correlated active-space calculations on a finite, non-periodic surface fragment. We demonstrate the workflow for the hydrogenation step RuH$_2$(N$_2$)* $\rightarrow$ RuH(NNH)* at an isolated Ru$_1$ site on Ru(0001). Starting from a periodically relaxed surface structure, a first-shell finite fragment is extracted and used to construct a correlated active-space description. We first use Active Atomic Valence Space (AVAS) to identify orbitals involved in the Ru-N/N-H bond reorganization, then truncate the space using natural orbitals to reduce computational cost while retaining the relevant correlated degrees of freedom. We map the reduced active-space Hamiltonian to qubits with the Jordan-Wigner (JW) transformation and solve it with the adaptive derivative-assembled pseudo-Trotter variational quantum eigensolver (ADAPT-VQE). Contributions from dynamic electron correlation beyond the selected active space are further examined using strongly contracted NEVPT2 and DSRG-MRPT2. For relaxed reaction structures, AVAS produces a 22-qubit active space, which natural-orbital truncation compresses to a 16-qubit representation. The 16-qubit truncated spaces reproduce the corresponding untruncated CASCI state energies to within 0.21 kcal/mol, with a reaction-energy error of only 0.2 kcal/mol. The statevector ADAPT-VQE in this reduced space converges for both reaction states under a fixed pool-gradient stopping criterion. Strongly contracted NEVPT2 yields anomalously large, state-imbalanced corrections for the finite Ru fragment, while DSRG-MRPT2 retains an endothermic reaction energy at its default flow parameter, although the predicted magnitude remains strongly dependent on the flow parameter.

\end{abstract}

\section{Introduction}

Ammonia is vital to global food security as the primary feedstock for nitrogen fertilizers and is increasingly being considered as a hydrogen-rich, carbon-free energy carrier for a low-carbon economy.\cite{erisman2008century,foster2018catalysts} More than 90\% of the ammonia produced today still relies on the century-old Haber-Bosch process, in which iron-based catalysts operate at 400-600~$^{\circ}$C and pressures exceeding 150~bar; this single process is estimated to consume 1-2\% of the global energy supply, with a correspondingly disproportionate share of industrial greenhouse-gas emissions.\cite{smith2020current}

Ruthenium-based catalysts are the most active established alternative to iron under substantially milder conditions.\cite{rosowski1997ruthenium} On extended Ru(0001) terraces, N$_2$ dissociation proceeds through B$_5$-type step-edge sites and has long been used as a model system for first-principles studies of the classical dissociative mechanism.\cite{dahl2000dissociative,honkala2005ammonia,logadottir2003ammonia} Reducing particle size below the $\approx$2~nm threshold required to expose B$_5$ sites eliminates this pathway; sub-nanometric and single-atom Ru sites, following the broader single-atom catalysis paradigm established for other late transition metals,\cite{qiao2011single,wang2018heterogeneous} have instead been shown, both computationally and via operando kinetics, to promote an associative mechanism, forming *NNH$_x$ intermediates via stepwise hydrogenation that circumvents direct N$\equiv$N cleavage.\cite{peng2024single,li2022size,qiu2019pure} Recently, an operando/DFT study identified RuH$_2$(N$_2$) as an intermediate state and the hydrogenation step of RuH$_2$(N$_2$)* $\rightarrow$ RuH(NNH)* as having the largest intrinsic barrier along the associative pathway.\cite{gupta2025mechanistic} Because this N-N/N-H bond-reorganization step is also particularly sensitive to known GGA-DFT limitations, the present work focuses on its reactant and product-state energetics.

Approximate Kohn-Sham DFT\cite{hohenberg1964inhomogeneous,kohn1965self} can be unreliable for this type of bond reorganization due to static correlation, self-interaction error, and near-degenerate electronic configurations, particularly in transition-metal chemistry relevant to heterogeneous catalysis.\cite{cohen2012challenges,verma2020status} Such limitations can become especially pronounced during bond-reorganization processes where $\pi$-backbonding and new $\sigma$-bond formation compete, as in N$_2$ activation. We therefore treat this step with a correlated method beyond a single exchange-correlation functional.

Multiconfigurational methods such as CASSCF/CASCI and NEVPT2 can capture static and dynamic correlation beyond single-reference DFT, but the exponential growth of the determinant space limits the active spaces accessible for realistic catalytic fragments. VQE\cite{peruzzo2014variational} and ADAPT-VQE offer a quantum alternative by representing the correlated wavefunction through parameterized quantum-state preparation, and have been applied to molecular ground- and excited-state electronic-structure problems on both simulators and early quantum hardware.\cite{tilly2022variational,cao2019quantum} Recent proof-of-concept studies have combined active-space selection or embedding with VQE for heterogeneous catalytic and materials problems, including O$_2$ dissociation on Pt-based surfaces,\cite{di2024platinum} defects in diamond,\cite{ma2020quantum} water dissociation on Mg,\cite{gujarati2023quantum} and alkene hydrogenation.\cite{li2022toward}

A realistic transition-metal active space places multireference and quantum electronic-structure methods in a computationally challenging regime. Dense orbital manifolds and near-degeneracies of this kind are a long-standing source of intruder-state divergences in perturbative treatments such as CASPT2, historically addressed through empirical real or imaginary level shifts\cite{forsberg1997multiconfiguration}; strongly contracted NEVPT2 was developed in part to avoid this failure mode through its Dyall-Hamiltonian-based zeroth-order partitioning,\cite{angeli2001introduction} and we use it here as the first dynamic-correlation correction. The same dense, near-degenerate orbital structure also complicates variational quantum algorithms, since convergence to the true ground state becomes harder to achieve as the density of near-degenerate configurations increases.\cite{bauer2020quantum} The problem is especially pronounced for finite metallic fragments, where inactive, active, and external orbitals are not cleanly separated as often in conventional closed-shell molecules.\cite{veryazov2011select} We therefore check each reduction step directly on the Ru fragment rather than assume that a protocol developed for molecular systems will behave the same way in a metallic cluster.

In this work, we connect periodic DFT relaxation of the Ru(0001) catalytic system to correlated calculations on a finite, non-periodic fragment, extending the AVAS/NO + ADAPT-VQE strategy previously demonstrated for Pt-based oxygen reduction\cite{di2024platinum} to the hydrogenation step RuH$_2$(N$_2$)* $\rightarrow$ RuH(NNH)*. A chemically motivated AVAS active space is constructed and systematically compressed using natural-orbital truncation before the resulting fermionic Hamiltonian is mapped onto qubits for ADAPT-VQE simulation, benchmarked directly against exact CASCI. NEVPT2 and DSRG-MRPT2 are examined as complementary approaches for recovering dynamic correlation outside the selected active space, with NEVPT2's stability tested directly on the finite Ru fragment and DSRG-MRPT2 evaluated across its flow parameter rather than only at a single default setting. All three methods show limitations, though of different kinds. NEVPT2 produces an anomalously large correction that reverses the reaction from endothermic to exothermic. ADAPT-VQE satisfies its gradient-norm stopping criterion but still leaves a non-negligible residual error for the reactant. DSRG-MRPT2 preserves the endothermic reaction character but remains quantitatively sensitive to its flow parameter. The calculations also let us check whether adding explicit static and dynamic correlation changes the reaction picture obtained previously with GGA-DFT.\cite{gupta2025mechanistic}

\section{Methodology}

\begin{figure}[htbp]
\centering
\includegraphics[width=\textwidth]{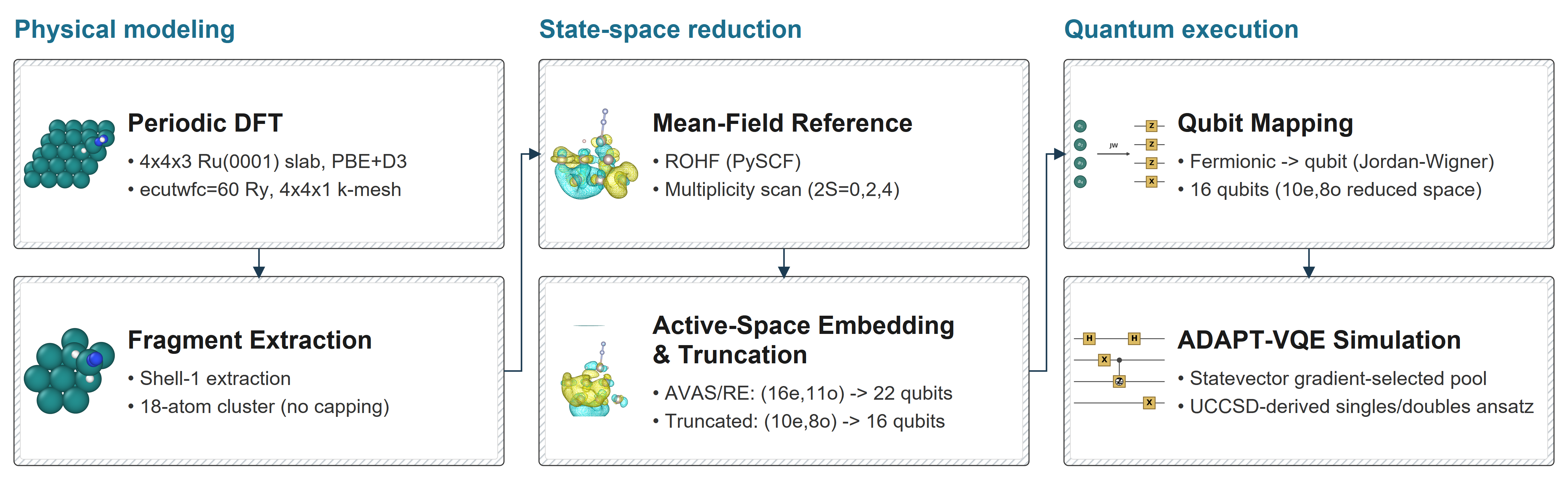}
\caption{Computational workflow. The six stages are grouped into three parts: physical modeling (periodic DFT relaxation, fragment extraction), state-space reduction (mean-field reference, active-space embedding and natural-orbital truncation), and quantum execution (fermionic-to-qubit mapping, ADAPT-VQE).}
\label{fig:pipelinearch}
\end{figure}

Fig.~\ref{fig:pipelinearch} summarizes the workflow through three interconnected parts: physical modeling, state-space reduction, and quantum execution. Periodic DFT is first used to relax the Ru(0001) catalytic structure, from which a finite, non-periodic fragment is extracted for correlated treatment. A mean-field reference is then constructed and used to define a chemically motivated active space, which is further reduced by orbital truncation to obtain a compact many-electron problem. The resulting fermionic Hamiltonian is mapped onto qubits and solved using ADAPT-VQE, with the quantum results benchmarked against classical active-space references.

\subsection{Reaction Model and Periodic DFT}
\label{sec:reactionmodel}
\label{sec:periodicdft}

\begin{figure}[htbp]
\centering
\includegraphics[width=1.00\textwidth]{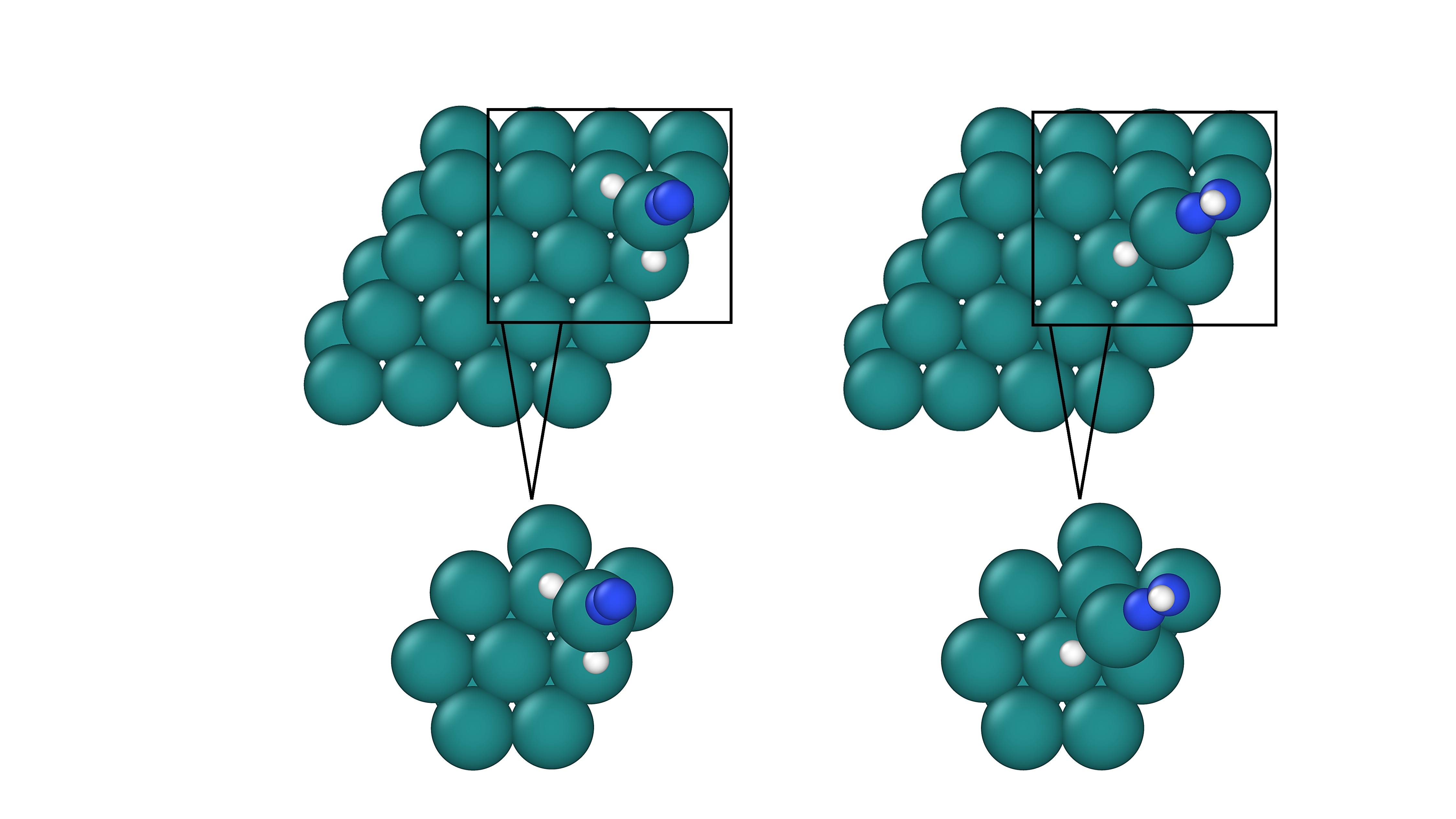}
\caption{The two configurations considered throughout this work, reactant (left) vs.\ product (right) in each row. Top: the converged periodic Ru(0001) slab region around the active site. Bottom: the finite, non-periodic 18-atom Ru$_{14}$N$_2$H$_2$ fragment extracted. Teal: Ru; blue: N; white: H.}
\label{fig:configs}
\end{figure}

The Ru(0001) surface was modeled using a close-packed hexagonal ($4\times4\times3$) slab containing 48 Ru atoms, with lattice parameters $a = 2.706$~\AA\ and $c = 4.282$~\AA, which matches the experimental value.\cite{kittel1976introduction} A vacuum region of 11~\AA\ was introduced along the surface-normal direction to minimize interactions between periodic images, while the bottom Ru layer was held fixed to represent the bulk-terminated substrate. A single Ru adatom is placed on a three-fold hollow site of the terrace to form the active Ru$_1$ site. The reactant state, RuH$_2$(N$_2$)*, consists of a coordinated N$_2$ molecule with two dissociatively adsorbed hydrogen atoms. The product state, RuH(NNH)*, is reached by migration of one hydrogen onto the terminal nitrogen, forming an associative N-NH fragment with one hydrogen remaining Ru-bound. The two states are isomeric: identical stoichiometry (one Ru adatom, two N atoms, and two H atoms) on the same slab. Fig.~\ref{fig:configs} (Section~\ref{sec:fragextraction}) shows both converged configurations. Because the two states have identical stoichiometry, the reaction energy is calculated directly as
\begin{equation}
\Delta E_{\mathrm{hydrog}} \;=\; E[\mathrm{RuH(NNH)^*}] \;-\; E[\mathrm{RuH_2(N_2)^*}]
\label{eq:dE}
\end{equation}

Total energies and atomic forces were calculated using plane-wave density functional theory (DFT) with ultrasoft pseudopotentials as implemented in Quantum ESPRESSO.\cite{giannozzi2009quantum,giannozzi2017advanced} Exchange and correlation were described using the Perdew-Burke-Ernzerhof (PBE) generalized-gradient approximation,\cite{perdew1996generalized} together with the semi-empirical Grimme-D3 dispersion correction.\cite{grimme2010consistent} Ruthenium was represented by a scalar-relativistic ultrasoft pseudopotential\cite{vanderbilt1990soft} that includes the $4s$ and $4p$ semicore states in the valence manifold ($Z_v=16$), while standard ultrasoft pseudopotentials were used for N and H. Metallic occupations were treated with Methfessel-Paxton smearing\cite{methfessel1989high} with a width of $\sigma=0.02$~Ry. The plane-wave cutoff and Monkhorst-Pack $k$-point mesh were converged on the bare Ru(0001) slab prior to production calculations (Fig.~\ref{fig:fig1}; Tables~\ref{tab:cutoffscan} and~\ref{tab:kpointscan}, SI). A wavefunction cutoff of ecutwfc=60~Ry (charge density cutoff, ecutrho=480~Ry) and a $4\times4\times1$ mesh were adopted for production. In the wavefunction-cutoff convergence scan, performed at fixed charge-density dual, the 60 Ry point differed from the tightest tested cutoff by approximately 22 meV/atom, while the adopted $4\times4\times1$ mesh differed by 0.46 meV/atom. Because the reactant and product have identical stoichiometry on the same Ru slab, residual cutoff errors are expected to largely cancel in $\Delta E_{\mathrm{hydrog}}$. Geometries were relaxed using the BFGS algorithm with a total-energy convergence threshold of $1\times10^{-5}$~Ry for both states, and a force convergence threshold of $5\times10^{-3}$~Ry/Bohr for both reactant and product.

\subsection{Fragment Extraction}
\label{sec:fragextraction}

From each relaxed periodic structure, a finite, non-periodic cluster is extracted for subsequent molecular quantum-chemistry calculations. The cluster comprises the N$_2$H$_2$ adsorbate, the Ru$_1$ adatom, and the first coordination shell ($n=1$) of nearest-neighbor Ru atoms within a radius of approximately $1.15a$, where $a=2.706$~\AA\ is the Ru(0001) lattice constant defined in Section~\ref{sec:reactionmodel}. This procedure yields an 18-atom fragment with the composition Ru$_{14}$N$_2$H$_2$ for the present reaction states, treated without capping atoms at the cluster boundary. The construction follows the general finite-cluster strategy used to describe localized surface reactions with correlated wavefunction methods\cite{gujarati2023quantum,lau2021regional,sun2016quantum} and is analogous to the Pt$_{19}$O$_2$ fragments employed for O$_2$ dissociation.\cite{di2024platinum}

For the subsequent molecular quantum-chemistry calculations, the extracted fragment is treated as an isolated molecular system. The present approach therefore constitutes finite-cluster extraction rather than explicit quantum embedding. The energetic effect of this extraction is assessed at the PBE-D3 level in Section~\ref{sec:relaxedgeom} by comparing the periodic slab with the corresponding extracted fragment. This check probes the reaction-energy sensitivity to fragmentation at the DFT level; possible boundary effects on the correlated orbital spectrum are considered separately in Section~\ref{sec:discussion}.

\subsection{Active-Space Construction}
\label{sec:activespace}

Mean-field and multireference calculations are performed using PySCF.\cite{sun2020recent,sun2018pyscf} An unrestricted Hartree-Fock (UHF) scan over nominal spin states $2S=0$, 2, and 4 is first performed for the neutral fragment, followed by restricted open-shell Hartree-Fock (ROHF) at the selected multiplicity. The UHF solutions are known to exhibit severe spin contamination,\cite{schlegel1986potential} with $\langle S^2\rangle$ values of approximately 15-23 compared to nominal values of 0, 2, and 6, and several calculations do not meet the SCF convergence criterion (Table~\ref{tab:spinscan}). The lowest-energy nominal $2S=0$ solution is therefore retained only as a practical orbital reference and should not be interpreted as establishing the physical ground-state spin. Accordingly, the correlated energies reported below should be interpreted as results within the imposed singlet sector rather than as a definitive determination of the physical spin ground state of the finite Ru fragment.

Ruthenium is described using the def2-SVP basis\cite{weigend2005balanced} with its associated effective core potential,\cite{andrae1990energy} while N and H are treated with the all-electron cc-pVDZ basis.\cite{dunning1989gaussian} Active orbitals are selected using the AVAS procedure,\cite{sayfutyarova2017automated} with projections onto Ru $4d/5s$, N $2s/2p$, and H $1s$ minimal-basis reference functions. The AVAS overlap threshold is scanned and the candidate spaces are validated by CASCI, following the threshold-selection strategy.\cite{di2024platinum} Plain AVAS is used for all production calculations, while the regional embedding construction\cite{di2024platinum,lau2021regional} is examined separately as a diagnostic for the NEVPT2 instability discussed in Section~\ref{sec:nevpt2results}.

For both reaction states, the selected AVAS threshold yields a (16e,11o) active space, corresponding to 22 spin-orbitals under the Jordan-Wigner mapping. The corresponding active-space Hamiltonian is
\begin{equation}
\hat H =
\sum_{pq} h_{pq}\hat a_p^{\dagger}\hat a_q
+
\frac{1}{2}
\sum_{pqrs}
h_{pqrs}
\hat a_p^{\dagger}\hat a_q^{\dagger}\hat a_r\hat a_s,
\label{eq:hamiltonian}
\end{equation}
where $h_{pq}$ and $h_{pqrs}$ are the one- and two-electron integrals in the active molecular-orbital basis, and $\hat a_p^{\dagger}$ and $\hat a_p$ are fermionic creation and annihilation operators.

\subsection{Natural-Orbital Active-Space Truncation}
\label{sec:notrunc}

The (16e,11o) AVAS space is further reduced to obtain an active space more suitable for ADAPT-VQE benchmarking and compact qubit representations. Natural orbitals are obtained by diagonalizing the CASCI one-particle reduced density matrix, with the corresponding occupation numbers ranging from 0 to 2.\cite{lowdin1955quantum} Orbitals with occupations closest to 0 or 2 contribute least to the static correlation represented by the CAS and are therefore candidates for removal, following the general rationale of natural-occupation-based truncation.\cite{sosa1989selection}

For a retained orbital count $n_{\mathrm{active}}$, each reaction state independently retains the $n_{\mathrm{active}}$ orbitals whose occupations lie farthest from either 0 or 2. Fixing the retained orbital count, rather than a common occupation threshold, ensures that the reactant and product are represented using the same number of active orbitals and hence the same qubit count. The frozen orbitals are incorporated through a dressed one-electron Hamiltonian in which their Coulomb and exchange contributions are folded into the effective core Hamiltonian and their energy contribution into an additive constant. Because these orbitals are removed from within an already-defined CAS, rather than from the conventional inactive orbital space, this procedure is treated as an approximate active-space truncation.

The reduced-space FCI energy is compared against the parent (16e,11o) CASCI energy separately for each reaction state. A truncation is accepted only when the energy deviation is below 1~kcal/mol. The retained orbital count is scanned, and the smallest common $n_{\mathrm{active}}$ satisfying this criterion for both states is selected for the subsequent quantum calculations (Section~\ref{sec:notruncresults}).

\subsection{Qubit Mapping and ADAPT-VQE}
\label{sec:methods_adaptvqe}

The active-space Hamiltonian is mapped to qubits using the Jordan-Wigner transformation.\cite{jordanwigner1928} The eight retained spatial orbitals give 16 spin-orbitals, represented one-to-one by 16 qubits under this encoding, which preserves the fermionic anticommutation relations by introducing the required parity strings as products of Pauli-$Z$ operators. All ADAPT-VQE results reported in this work are obtained on this 16-qubit Jordan-Wigner Hamiltonian.

The Variational Quantum Eigensolver\cite{peruzzo2014variational} prepares a parameterized trial state on a quantum device and minimizes its energy expectation value via a classical optimizer. By the variational principle, the optimized energy remains an upper bound to the exact ground-state energy.\cite{tilly2022variational} ADAPT-VQE\cite{grimsley2019adaptive} replaces a fixed-form ans\"atz with one constructed adaptively: an ordered product of parameterized fermionic excitation operators, selected one at a time rather than fixed a priori,
\begin{equation}
|\psi(\boldsymbol{\theta})\rangle \;=\; \prod_i \exp(\theta_i \hat A_i)\, |\psi_{\mathrm{ref}}\rangle
\label{eq:ansatz}
\end{equation}
Each $\hat A_i$ is drawn from a pool of single- and double-excitation operators (derived from UCCSD\cite{romero2019strategies}). At each iteration, the operator with the largest energy gradient is added to the ans\"atz until the pool gradient norm falls below a convergence threshold. A hardware-efficient qubit-excitation variant of this algorithm\cite{tang2021qubit} further reduces the circuit depth by directly targeting Pauli-string operators rather than fermionic excitations; the present work uses the original fermionic-operator-pool formulation. We use the adaptive ans\"atz to avoid introducing the full UCCSD operator set from the outset and to keep the variational circuit more compact, following the previous approach,\cite{di2024platinum} which reported a similarly sized active space. The parameters are classically optimized (BFGS, PennyLane\cite{bergholm2018pennylane}) against the variational cost function
\begin{equation}
E(\boldsymbol{\theta}) \;=\; \langle \psi(\boldsymbol{\theta}) | \hat H | \psi(\boldsymbol{\theta}) \rangle \;\geq\; E_0
\label{eq:evar}
\end{equation}
The ans\"atz is evaluated on a noiseless statevector simulator and benchmarked directly against the corresponding CASCI/FCI active-space reference.

\subsection{NEVPT2 and DSRG-MRPT2 Correction}
\label{sec:dsrgmethod}

Dynamic correlation outside the active space is first examined using strongly contracted $n$-electron valence state perturbation theory (NEVPT2),\cite{angeli2001introduction,guo2016n} applied to the CASCI active-space reference as a second-order correction. The correction can be written schematically as
\begin{equation}
E^{(2)} \;=\; -\sum_k \frac{|\langle \Psi_{\mathrm{CAS}} | \hat H | \Phi_k \rangle|^2}{E_k - E_{\mathrm{CAS}}}
\label{eq:nevpt2}
\end{equation}
The corrected total energy, $E_{\mathrm{total}} = E_{\mathrm{active\text{-}space}} + E^{(2)}$, is used for both reactant and product states when evaluating $\Delta E_{\mathrm{hydrog}}$ in Eq.~\eqref{eq:dE}. A strongly contracted NEVPT2 correction was evaluated as a test of dynamic correlation outside the active space. On the present finite Ru fragment, the correction becomes anomalously large and strongly state-imbalanced, reversing the reaction from endothermic to exothermic. Because the NEVPT2 correction is unstable for this fragment, we also tested DSRG-MRPT2.\cite{evangelista2014driven,li2015multireference} DSRG-MRPT2 suppresses contributions associated with small energy gaps through a flow parameter dependent regularization, providing a numerically smoother treatment of near-degenerate contributions. The DSRG regularization can be represented schematically as
\begin{equation}
E^{(2)}_{\mathrm{DSRG}} \;=\; -\sum_k \frac{|\langle \Psi_{\mathrm{CAS}} | \hat H | \Phi_k \rangle|^2 \left(1 - e^{-s\Delta_k^2}\right)}{\Delta_k}, \qquad \Delta_k = E_k - E_{\mathrm{CAS}},
\label{eq:dsrg}
\end{equation}
where $s$ is a flow parameter that controls the strength of the regularization. $E^{(2)}_{\mathrm{DSRG}}$ remains finite in the $\Delta_k \to 0$ limit for any $s > 0$, rather than requiring a post-hoc level shift as in level-shifted CASPT2.\cite{andersson1992second} We therefore test DSRG-MRPT2 on the same finite-fragment active space.

The DSRG-MRPT2 calculation is implemented via Forte,\cite{evangelista2024forte} interfaced to Psi4 through an FCIDUMP file spanning the regional core, active space, and the $n_{\mathrm{virt}}$ lowest-energy virtual orbitals. This is built from the same regional-core-dressed one- and two-electron Hamiltonian already used for the regional NEVPT2 treatment of Section~\ref{sec:nevpt2results}, independently reverified here to reproduce the full-core CASCI energy. A consolidated list of the production parameters used throughout this work is provided in Table~\ref{tab:repro} (SI).

\section{Results and Discussion}
\label{sec:results}

\subsection{DFT Energies}
\label{sec:relaxedgeom}

The optimized structures are shown in Fig.~\ref{fig:configs}. In the converged reactant structure, the N$_2$ molecule remains largely intact, with an N-N bond length of 1.13~\AA\ and a Ru-N distance of 1.97~\AA\ to the proximal nitrogen. The N-N bond is only slightly longer than that of gas-phase N$_2$ (1.10~\AA\cite{huber2013molecular}), indicating predominantly molecular end-on adsorption. After one hydrogen atom migrates to the terminal nitrogen, the product shows a shorter Ru-N bond of 1.79~\AA\ and an elongated N-N bond of 1.22~\AA, together with formation of a 1.04~\AA\ N-H bond. These structural changes are consistent with a stronger Ru-N interaction and progressive weakening of the N-N bond as N$_2$ becomes hydrogenated.\cite{fryzuk2000continuing} The periodic DFT calculation gives $\Delta E_{\mathrm{hydrog}}(\mathrm{DFT})=+27.5$~kcal/mol ($+1.19$~eV), so hydrogen transfer is endothermic. The corresponding N$_2$H$_2 \rightarrow$ NNH$+$H step in our earlier Ru-based free-energy profile rises by approximately $+1.13$~eV.\cite{gupta2025mechanistic} The two numbers are not directly equivalent, since they come from different models and thermodynamic quantities, but they place the hydrogenation step on a similar energy scale. These relaxed geometries are subsequently used to construct the finite fragments for the active-space calculations of Section~\ref{sec:notruncresults}.

To assess whether finite-fragment extraction itself produces a large change in the reaction energetics, $\Delta E_{\mathrm{hydrog}}$ was additionally evaluated at the same PBE-D3 level used for the periodic calculation. The extracted 18-atom Ru$_{14}$N$_2$H$_2$ structures were placed in a $30\times30\times30$~\AA\ cubic supercell and sampled at the $\Gamma$ point, with the large vacuum separation used to suppress interactions between periodic images, using the same parameters as in the slab calculation. No further structural relaxation was performed; the coordinates were those extracted directly from the converged periodic structures. This matched fragment calculation gives $\Delta E_{\mathrm{hydrog}}^{\mathrm{fragment,DFT}}=+25.4$~kcal/mol, only $2.1$~kcal/mol below the periodic-slab value. Thus, at the PBE-D3 level, truncation to the first-shell fragment introduces only a modest change in reaction energy. For comparison, the finite-fragment CASCI reaction energy reported in Section~\ref{sec:notruncresults} is $+148.2$~kcal/mol. The much larger difference between the fragment PBE-D3 and CASCI results therefore cannot be explained by the PBE-D3-level fragmentation shift alone. This comparison does not, however, establish fragment-size convergence of the correlated Hamiltonian or of the dense Ru orbital spectrum, which remains a separate question considered below.

\subsection{Converged-Geometry Active Space and Natural-Orbital Truncation}
\label{sec:notruncresults}

AVAS applied to the converged-geometry fragment (18 atoms, 472 contracted basis functions) at threshold 0.905 selects an equal-dimensional, state-specific (16e, 11o) active space for both reaction states (22 spin-orbitals under Jordan-Wigner), with CASCI energies of $-1426.20333803$~Ha (reactant) and $-1425.96719340$~Ha (product).

\begin{figure}[htbp]
\centering
\includegraphics[width=0.65\textwidth]{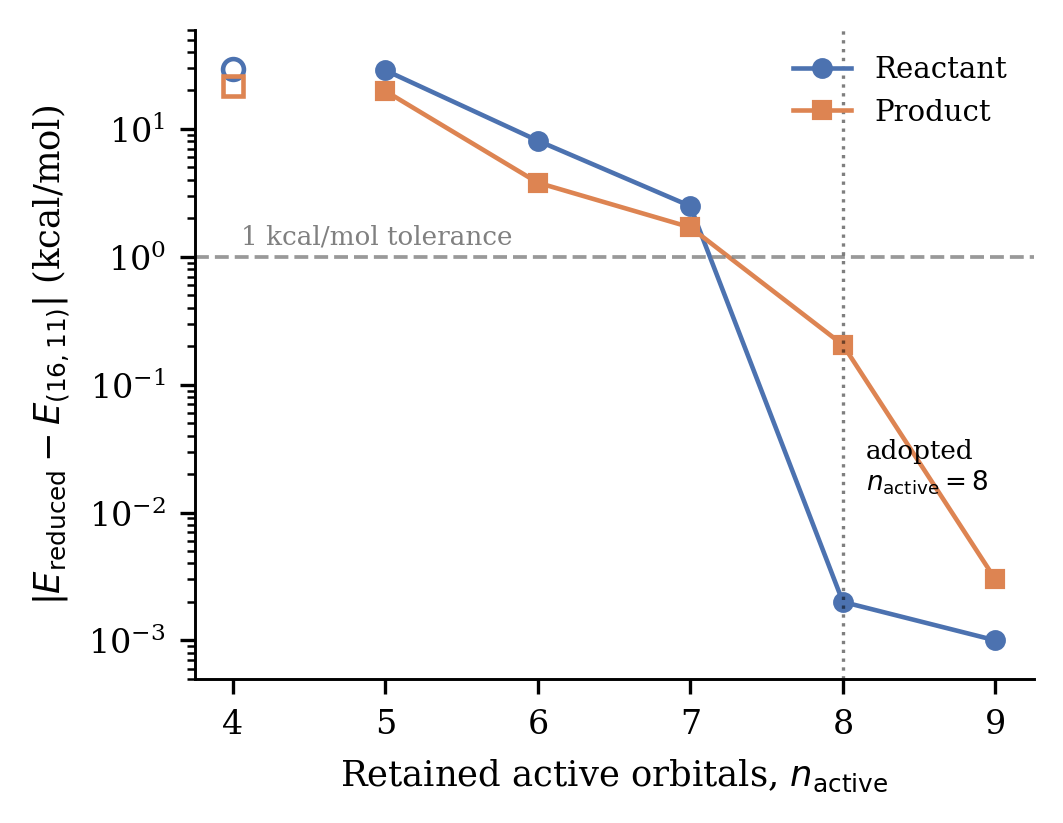}
\caption{Natural-orbital active-space truncation error vs.\ retained active-orbital count $n_{\mathrm{active}}$, relative to the full (16e,11o) CASCI reference, for both reaction states (log scale). Open markers at $n_{\mathrm{active}}=4$ denote the rejected reduction. The error drops sharply between $n_{\mathrm{active}}=7$ and 8, the adopted value (dotted line): by about three orders of magnitude for the reactant and about one order of magnitude for the product.}
\label{fig:fig4}
\end{figure}

Applying the natural-orbital truncation of Section~\ref{sec:notrunc}, we scanned the retained-orbital count $n_{\mathrm{active}}$ from 4 to 9 (Table~\ref{tab:notrunc}, SI; Fig.~\ref{fig:fig4}). The truncation error falls sharply between $n_{\mathrm{active}} = 7$ and $8$, from $+2.5$ to $+0.002$~kcal/mol for the reactant and from $+1.7$ to $+0.205$~kcal/mol for the product, and the occupation spectrum explains why: the three orbitals removed at \(n_\mathrm{active}=8\) have occupations of approximately 2.0, 2.0, and 1.9998, effectively doubly occupied within the parent CASCI description. The fully paired orbitals contribute negligible static correlation, whereas at $n_{\mathrm{active}}=7$ the additionally frozen orbital carries a meaningful fractional occupation. The aggressive reduction to $n_{\mathrm{active}}=4$ was tested first and rejected: its errors ($+29.2$~kcal/mol for the reactant and $+21.9$~kcal/mol for the product) far exceed the 1~kcal/mol tolerance.

\begin{figure}[htbp]
\centering
\includegraphics[width=1.00\textwidth]{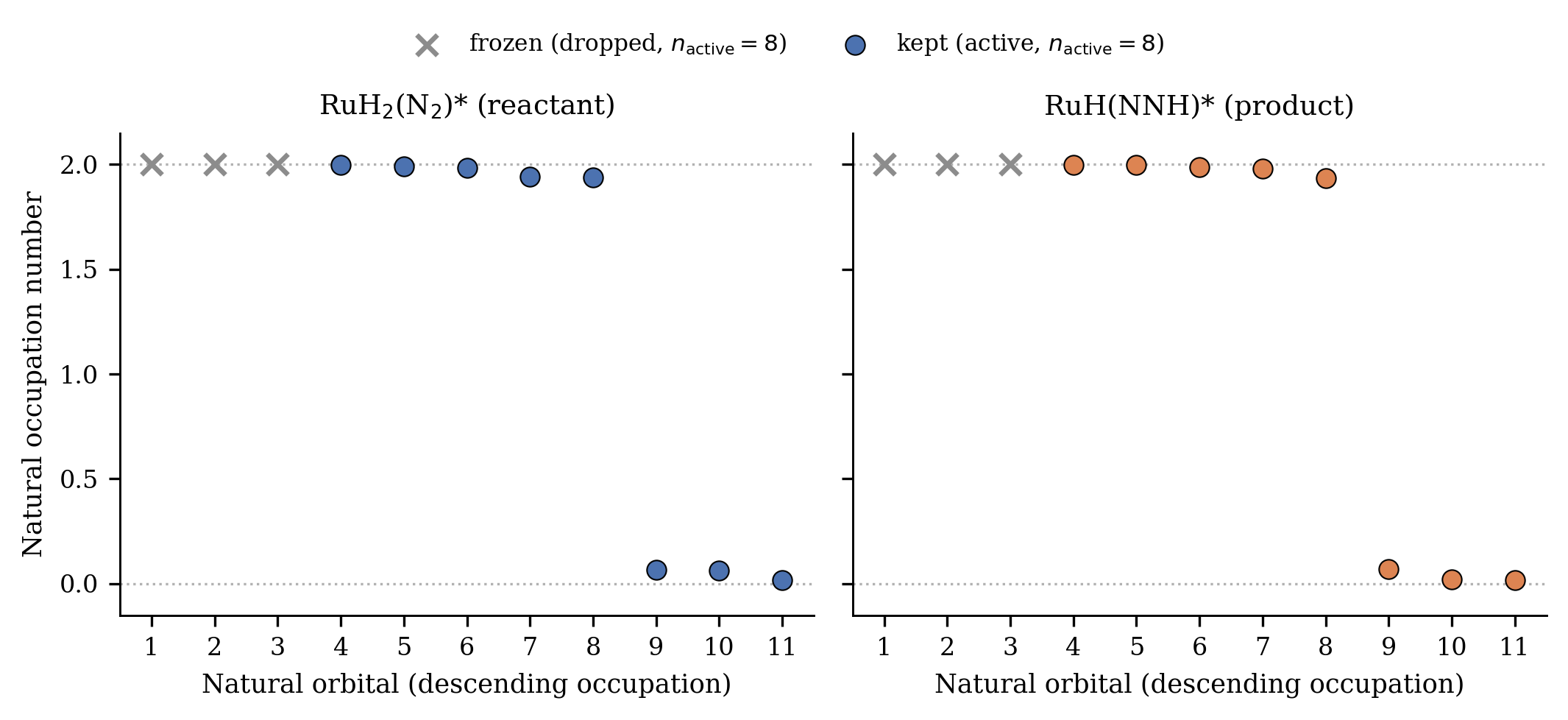}
\caption{Natural-orbital occupation-number spectrum of the (16e,11o) CASCI active space, both reaction states, sorted in descending order. The three frozen orbitals (crosses) sit within $2\times10^{-4}$ of a fully doubly occupied integer occupation for both states; the eight retained orbitals (filled markers, $n_{\mathrm{active}}=8$) span a continuous range down to $\sim$0.02, reflecting fractional/multireference character.}
\label{fig:fig5}
\end{figure}

Fig.~\ref{fig:fig5} shows the full natural-occupation spectrum behind this sharp drop (Table~\ref{tab:naturalorb}, SI): for both states, the three orbitals dropped at $n_{\mathrm{active}}=8$ are clearly separated from the eight retained ones, sitting within $2\times10^{-4}$ of a fully doubly occupied integer occupation, whereas the eight retained orbitals span a continuous range of occupations from near 2 down to approximately 0.02.

We adopt $n_{\mathrm{active}}=8$ for the converged-geometry active space: an equal-dimensional, state-specific (10e, 8o) space, 16 spin-orbitals/16 qubits under Jordan-Wigner, reproducing the original (16e,11o) CASCI state energies to within 0.002~kcal/mol (reactant) and 0.205~kcal/mol (product). At the level of reaction energy that these two state energies combine into, the truncation error is smaller still: the reduced-space FCI reaction energy ($+148.4$~kcal/mol) differs from the untruncated (16e,11o) CASCI reaction energy ($+148.2$~kcal/mol) by $0.2$~kcal/mol, since the two per-state errors partially cancel rather than add. Mapped via Jordan-Wigner, the full (16e,11o)/22-qubit Hamiltonian has 20,890 (reactant) / 20,930 (product) Pauli terms, and the reduced (10e,8o)/16-qubit Hamiltonian has 5,789 / 5,793, a $\sim$3.6-fold reduction in term count from a 6-qubit reduction that closely tracks the $(22/16)^4 \approx 3.57$ scaling expected for Jordan-Wigner-mapped two-electron integrals in spin-orbital count (Fig.~\ref{fig:fig6}, SI).

\begin{figure}[htbp]
\centering
\begin{minipage}{0.50\textwidth}
\centering
\includegraphics[width=\textwidth]{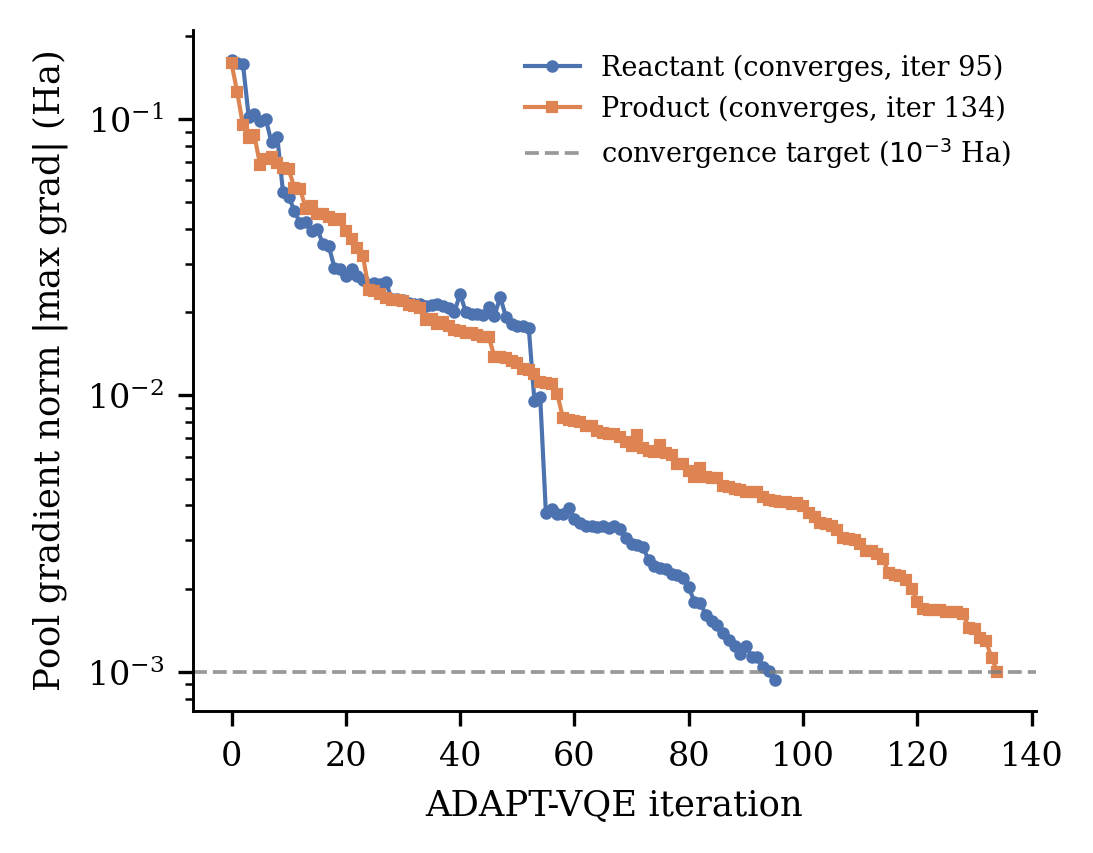}
\centerline{(a)}
\end{minipage}\hfill
\begin{minipage}{0.50\textwidth}
\centering
\includegraphics[width=\textwidth]{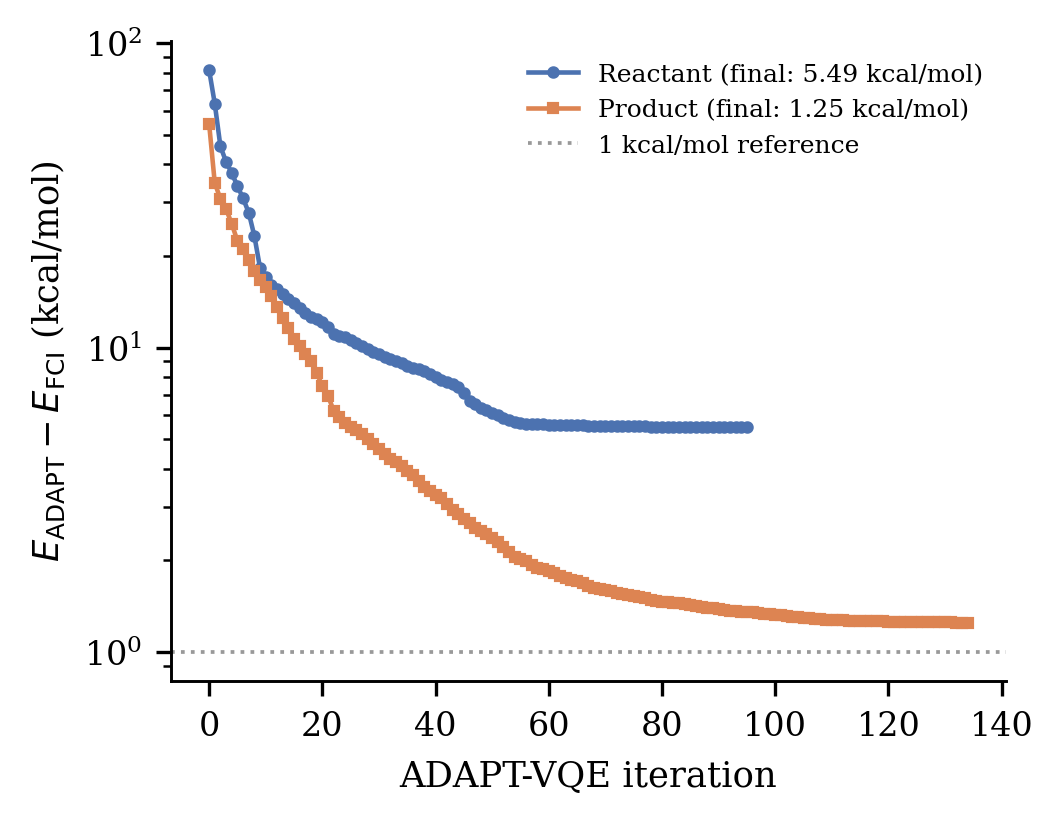}
\centerline{(b)}
\end{minipage}
\caption{ADAPT-VQE convergence diagnostics for the natural-orbital-truncated (10e,8o), 16-qubit active space. (a) Pool gradient norm versus iteration for the reactant and product, with the $1\times10^{-3}$~Ha convergence threshold indicated by the dashed line.  (b) Variational energy error $E_{\mathrm{ADAPT}}-E_{\mathrm{FCI}}$ versus iteration, in kcal/mol, relative to the corresponding reduced-space FCI reference; the dotted horizontal line marks the 1~kcal/mol tolerance.}
\label{fig:convergence}
\end{figure}

Statevector ADAPT-VQE calculations were performed on the reduced (10e, 8o) active space for both reaction states, using the spin-conserving singles/doubles operator pool described in Section~\ref{sec:methods_adaptvqe}. The pool contained 315 candidate excitations (30 singles and 285 doubles), and convergence was defined by a gradient-norm threshold of $1\times10^{-3}$~Ha. The reactant converged after 95 iterations with 96 operators in the final ans\"atz and a terminal gradient norm of $9.32\times10^{-4}$, yielding an energy of $-1426.19459147$~Ha, 5.49~kcal/mol above the corresponding reduced-space FCI reference ($-1426.20333468$~Ha). The product converged after 134 iterations with 135 operators and a terminal gradient norm of $9.99\times10^{-4}$, reaching $-1425.96487891$~Ha, 1.25~kcal/mol above its FCI reference ($-1425.96686694$~Ha) (Fig.~\ref{fig:convergence}b).

Combining the reactant and product state energies gives an ADAPT-VQE reaction energy of $\Delta E_{\mathrm{hydrog}}=+144.1$~kcal/mol, compared with a reduced-space FCI reference of $+148.4$~kcal/mol, consistent with the untruncated (16e,11o) CASCI value of $+148.2$~kcal/mol; the resulting reaction-energy deviation is $4.2$~kcal/mol. Both reaction states satisfy the same $10^{-3}$~Ha gradient threshold, but this criterion does not give comparable energy accuracy for the two: the reactant remains 5.49~kcal/mol above FCI, whereas the product error is only 1.25~kcal/mol, so most of the 4.2~kcal/mol reaction-energy error originates from the reactant. This behavior resembles the false-gradient trough reported for ADAPT-VQE,\cite{grimsley2019adaptive} where the pool gradient becomes small before the wavefunction is sufficiently close to the ground state. The gradient threshold should therefore be interpreted only as a convergence criterion, rather than as an indicator of comparable energy accuracy between the two states. Testing tighter thresholds and alternative operator pools would help distinguish these effects.

The present ADAPT-VQE setup was not specifically tailored to this kind of metallic, near-degenerate orbital manifold: it uses a generic UCCSD-derived singles/doubles pool\cite{romero2019strategies} at a fixed 16-qubit natural-orbital-truncated active space. A more targeted excitation pool may be more effective than tightening the threshold alone, although we have not tested that here; retaining additional active orbitals, at a higher qubit cost, could similarly restore a cleaner active separation and reduce the residual error. All ADAPT-VQE results reported here were obtained on a noiseless statevector simulator, so it is unknown whether the same convergence behavior holds under realistic NISQ noise and error mitigation.\cite{bauer2020quantum} Section~\ref{sec:hwcompare} (SI) briefly provides a compilation-level estimate of this hardware gap: transpiling the converged ADAPT-VQE ans\"atze to representative contemporary IBM and IonQ targets substantially increases circuit depth and total gate count relative to the
corresponding logical circuits.

\FloatBarrier

\subsection{NEVPT2 and DSRG-MRPT2 Dynamic Correction}
\label{sec:nevpt2results}

\begin{figure}[htbp]
\centering
\includegraphics[width=1.00\textwidth]{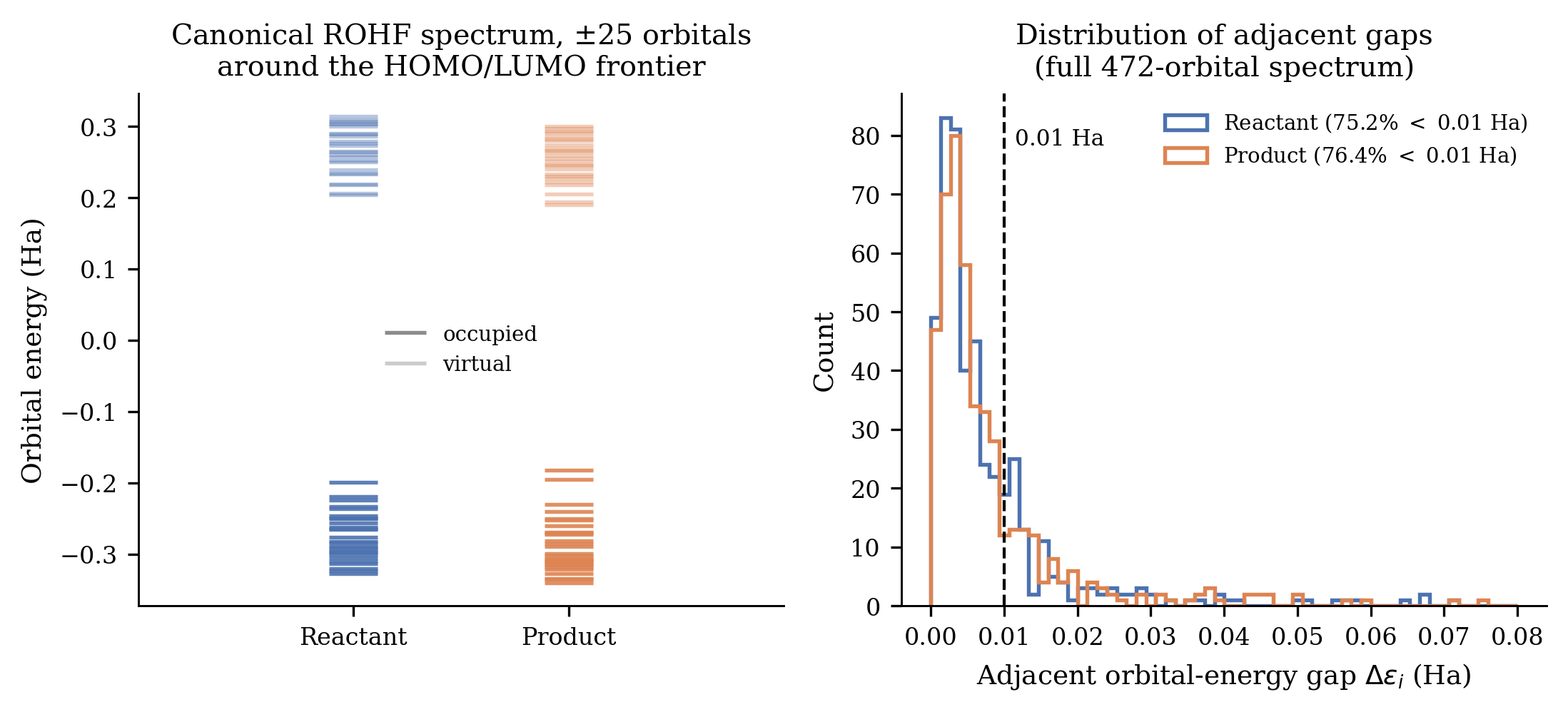}
\caption{Canonical ROHF orbital-energy structure underlying the AVAS construction, finite Ru fragment, both reaction states. Left: orbital energies within 25 orbitals of the HOMO/LUMO frontier (dark: occupied, light: virtual); the active space selected by AVAS is a rotated subspace concentrated at this frontier rather than a contiguous block in energy order. Right: distribution of adjacent orbital-energy gaps across the full 472-orbital spectrum, with 75.2\% (reactant) and 76.4\% (product) falling below the 0.01~Ha threshold (dashed line), consistent with the near-continuous, quasi-metallic manifold discussed in the text.}
\label{fig:orbspectrum}
\end{figure}

A direct strongly contracted NEVPT2 correction (Eq.~\eqref{eq:nevpt2}) was attempted on top of the (16e,11o) CASCI active-space reference of Section~\ref{sec:notruncresults} but produces an anomalously large correction that reverses the reaction from endothermic to exothermic. The canonical ROHF spectrum is dense over much of the orbital manifold: 75.2\% of adjacent orbital-energy gaps for the reactant and 76.4\% for the product fall below 0.01~Ha across the full 472-orbital spectrum (Fig.~\ref{fig:orbspectrum}), consistent with the Ru effective core potential already stripping true atomic core character. This dense spacing may contribute to the instability of the perturbative correction. However, strongly contracted NEVPT2's Dyall-Hamiltonian-based zeroth-order partitioning is specifically constructed to avoid the intruder-state divergences that affect Fock-based methods such as CASPT2,\cite{angeli2001introduction} and we have not inspected NEVPT2's own internally contracted denominators directly. For the present fragment, we therefore report the NEVPT2 result as numerically unstable without assigning a specific microscopic mechanism. Restricting dynamic correlation to a spatially local ``regional'' core subset (Ru$_1$ adatom + adsorbate Mulliken population above a threshold), following the two-step AVAS/regional embedding NEVPT2 treatment,\cite{di2024platinum} improved single-state behavior but not the reaction energy: at the Mulliken threshold of 0.03, this gives $E^{(2)} = -1.489$~Ha (reactant) and $-2.705$~Ha (product) (perturber-term breakdown in Table~\ref{tab:nevpt2terms}, SI), corresponding to $\Delta E_{\mathrm{hydrog}}^{\mathrm{NEVPT2}} = -614.6$~kcal/mol against the $+148.2$~kcal/mol CASCI reference (Section~\ref{sec:notruncresults}), a reversal from endothermic to strongly exothermic, with a magnitude far larger than the CASCI reaction energy.

\begin{figure}[htbp]
\centering
\includegraphics[width=0.50\textwidth]{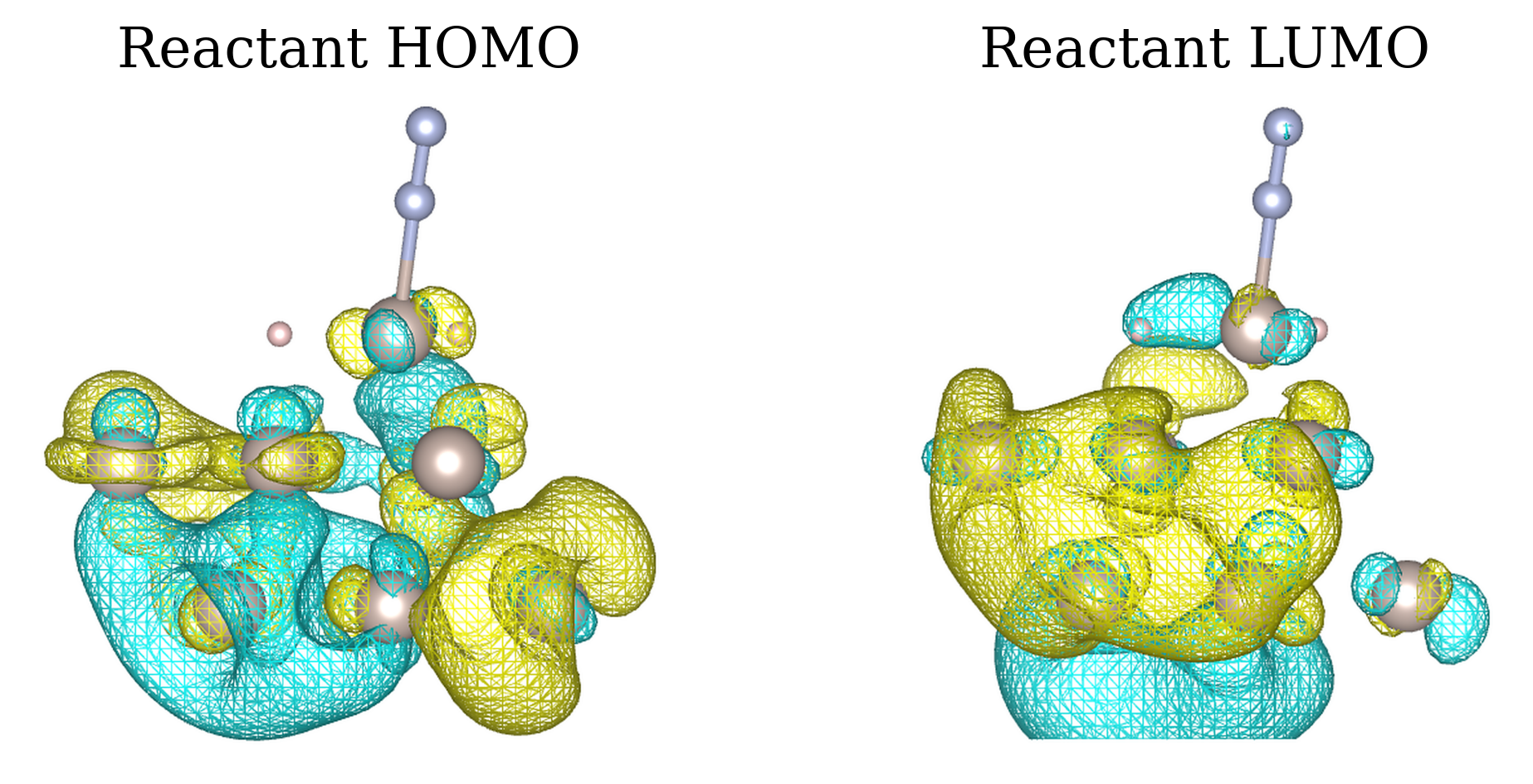}\\[6pt]
\includegraphics[width=0.50\textwidth]{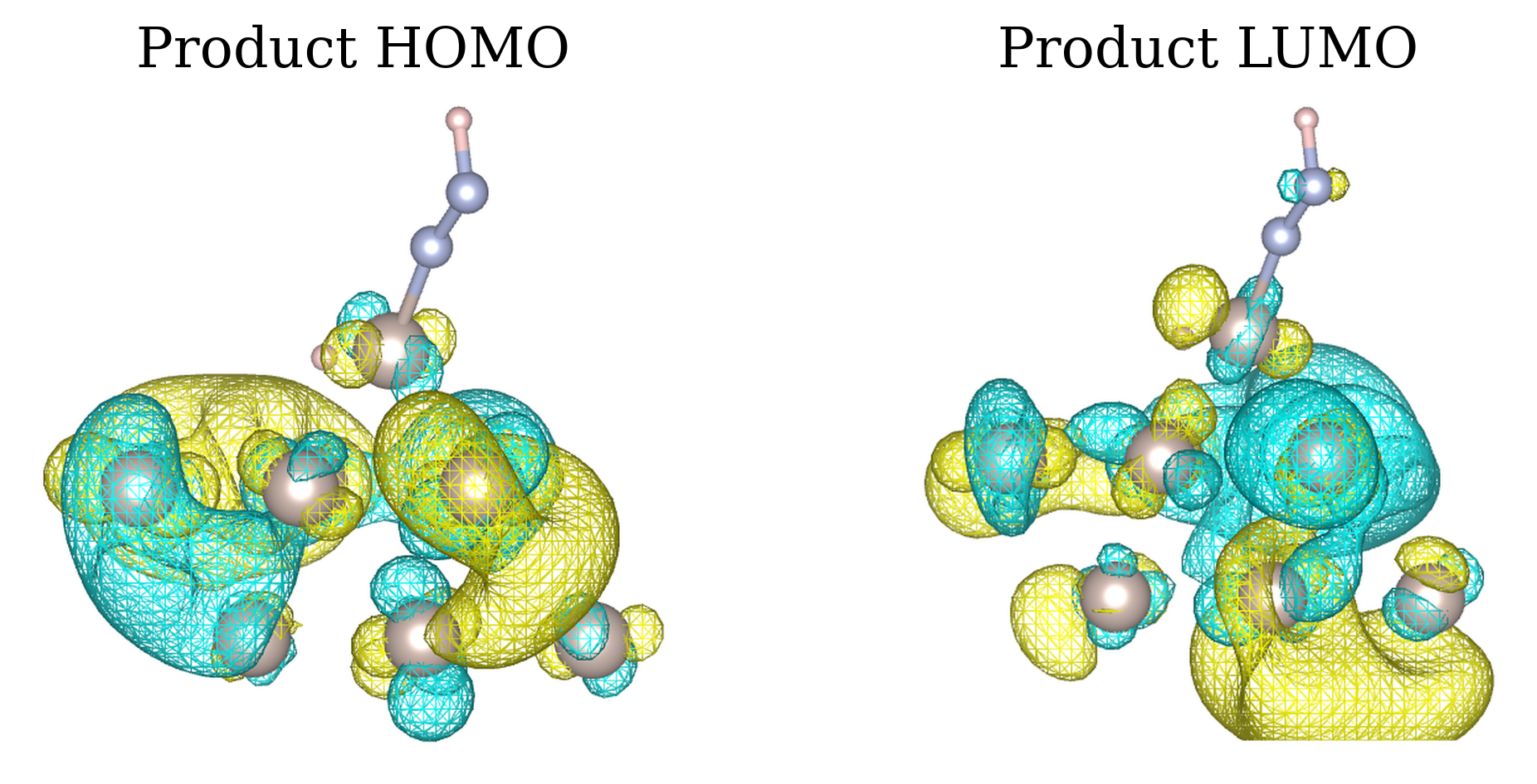}
\caption{Canonical ROHF HOMO and LUMO isosurfaces, reactant (top) vs.\ product (bottom), both relaxed geometries. Yellow/cyan denote opposite phase. In both states, both orbitals are concentrated on the Ru cluster rather than the N$_2$H$_2$ adsorbate.}
\label{fig:homolumo}
\end{figure}

As a targeted diagnostic for whether this instability traces to the AVAS projector itself, we repeated the active-space construction with the regional embedding variant,\cite{lau2021regional} which builds the virtual-orbital projectors from the full computational basis set rather than a minimal-basis representation, with the rest of the protocol unchanged. CASCI on the resulting active space validates the construction, agreeing with the plain-AVAS reaction energy to within $0.1$~kcal/mol ($+148.1$ vs.\ $+148.2$~kcal/mol). Applying the same regional-core NEVPT2 treatment on top of this space did not resolve the instability: $E^{(2)} = -1.408$~Ha (reactant) and $-2.668$~Ha (product), giving $\Delta E_{\mathrm{hydrog}}^{\mathrm{NEVPT2}} = -642.7$~kcal/mol, same behavior as the plain-AVAS result above. So the instability is not an artifact of the minimal-basis AVAS projector choice. Fig.~\ref{fig:homolumo} shows the canonical ROHF HOMO and LUMO of the full fragment (not the AVAS-rotated active-space orbitals) for both states. In both states and both orbitals, the isosurfaces are concentrated on the Ru cluster itself rather than the N$_2$H$_2$ adsorbate, consistent with the quasi-metallic orbital structure as a possible contributor to the NEVPT2 instability. It remains unclear how much of the dense spectrum is intrinsic to Ru and how much comes from the finite-cluster boundary. A fragment-size study would separate these contributions. If the instability persists, a more explicit embedding treatment, such as density matrix embedding theory,\cite{knizia2012density} or a different multireference treatment of dynamic correlation, may ultimately be required.

Unlike NEVPT2, DSRG-MRPT2 retains an endothermic reaction energy. Using the same regional-core treatment, DSRG-MRPT2's formally regularized denominator keeps the correction numerically well-behaved: at its literature-default parameter point ($n_{\mathrm{virt}}=50$, $s=0.5$\cite{li2015multireference,evangelista2024forte}), it gives $\Delta E_{\mathrm{hydrog}}^{\mathrm{DSRG}} = +70.5$~kcal/mol, matching the endothermic character of both the finite-fragment CASCI and periodic-DFT reaction energies. This agreement should not be interpreted as numerical convergence, however: the CASCI/DSRG-MRPT2 values are finite-fragment results on a Hamiltonian and system representation different from the periodic-DFT calculation, so the numerical proximity of the DSRG value to the DFT value does not indicate a converging hierarchy. The value is not quantitatively converged with respect to the flow parameter. A flow-parameter sensitivity test over three points spanning the typical literature range (0.3-1.0~Ha$^{-2}$,\cite{li2015multireference} fixed $n_{\mathrm{virt}}=50$) finds that the reaction energy varies from $+92.6$ to $+36.3$~kcal/mol across that range (Table~\ref{tab:dsrgsens}, SI), a $\sim$56~kcal/mol swing; with only three points tested, this indicates strong sensitivity rather than tracing the shape of a converged curve. Because the external virtual space is itself truncated at fixed $n_{\mathrm{virt}}=50$, this sensitivity to $s$ cannot yet be separated from the effect of virtual-space truncation; a joint convergence study in both $s$ and $n_{\mathrm{virt}}$ is therefore required before a quantitative DSRG-MRPT2 reaction energy can be assigned.

\label{sec:discussion}
Together, these calculations highlight a common challenge: the finite Ru cluster has a dense orbital spectrum with no clean separation between inactive, active, and external orbitals. For the perturbative methods, this makes the treatment of dynamic correlation especially sensitive to how the orbital space is partitioned and to the size of the energy denominators. The direct PBE-D3 fragment check of Section~\ref{sec:relaxedgeom} provides an important constraint on this interpretation: truncation from the periodic slab to the first-shell fragment changes the PBE-D3 reaction energy by only $2.1$~kcal/mol. Thus, the large difference between the PBE-D3 and CASCI reaction energies cannot be explained by the DFT-level fragmentation shift alone. This result does not, however, determine whether the near-continuous orbital spectrum or the instability of the subsequent multireference corrections is sensitive to the finite cluster boundary. Fragment-size convergence therefore remains necessary to separate an intrinsic metallic-spectrum effect from a finite-fragment artifact.

\section{Conclusion}

We combine periodic DFT with correlated active-space calculations on a finite, non-periodic surface fragment to study N$_2$ hydrogenation at a Ru single-atom site on Ru(0001). Periodic PBE-D3 relaxation of the RuH$_2$(N$_2$)* $\rightarrow$ RuH(NNH)* step gives a mildly endothermic reaction energy of $+27.5$~kcal/mol and provides the geometries used for the subsequent correlated electronic-structure calculations. A matched PBE-D3 calculation on the extracted fragment gives $+25.4$~kcal/mol, differing from the periodic value by only $2.1$~kcal/mol and indicating modest sensitivity to fragmentation at the DFT level.

AVAS followed by natural-orbital truncation reduces the active space from (16e, 11o) to (10e, 8o) while changing the CASCI reaction energy by only 0.2~kcal/mol. ADAPT-VQE reproduces the reduced-space FCI reaction energy to within 4.2~kcal/mol. However, the unequal residual errors of 5.49~kcal/mol for the reactant and 1.25~kcal/mol for the product, consistent with the false-gradient-trough behavior reported for ADAPT-VQE, show that satisfying the gradient criterion does not imply uniform energy accuracy across the two states. NEVPT2 becomes qualitatively unreliable for the finite Ru fragment, reversing the reaction from endothermic to exothermic, whereas DSRG-MRPT2 retains the endothermic character obtained with periodic DFT and finite-fragment CASCI but remains strongly dependent on the flow parameter. The principal remaining limitation is the quantitative treatment of dynamic correlation in the finite metallic fragment. Although the DFT-level fragmentation test indicates only a small energetic effect, fragment-size convergence is still required to determine whether the dense orbital spectrum and the behavior of the correlated methods depend on the finite-cluster boundary. A more explicit treatment of the metallic environment, convergence of the external virtual space, and broader benchmarking of multireference correlation methods are therefore needed.

\section*{Acknowledgment}
The author acknowledges Dr. Nagendra Nagaraja (CEO, QpiAI), Lakshya Priyadarshi (VP of Quantum Platforms and Solutions, QpiAI), Ashwanth Krishnan (VP of Software, QpiAI), and Dr. Ankit Patidar (Scientific Product Manager, QpiAI), and the management of QpiAI India Pvt. Ltd. for their support in providing the computational resources required for this work.

\section*{Competing Interests}
The author declares no competing financial or non-financial interests.

\bibliographystyle{achemso}
\bibliography{references}

\clearpage
\renewcommand{\thesection}{S\arabic{section}}
\setcounter{section}{0}
\renewcommand{\thetable}{S\arabic{table}}
\setcounter{table}{0}
\renewcommand{\thefigure}{S\arabic{figure}}
\setcounter{figure}{0}

\begin{center}
{\Large\bfseries Supporting Information}
\end{center}
\vspace{0.5em}

\noindent This Supporting Information accompanies the main text.

\section{Converged Cartesian Coordinates}
\label{si:coords}

\subsection{Full periodic slab (53 atoms: 49 Ru + N$_2$H$_2$)}

Cartesian Coordinates in \AA, as output by the converged BFGS relaxation.

\subsubsection*{Reactant, RuH$_2$(N$_2$)*}
\begin{verbatim}
Lattice vectors (Angstrom):
  a1 = 10.82400017   0.00000000   0.00000000
  a2 =  5.41200009   9.37385475   0.00000000
  a3 =  0.00000000   0.00000000  19.98312841

Ru     0.00000000     0.00000000     0.00000000
Ru     2.70600000     0.00000000     0.00000000
Ru     5.41200000     0.00000000     0.00000000
Ru     8.11800000     0.00000000     0.00000000
Ru     1.35300000     2.34346474     0.00000000
Ru     4.05900000     2.34346474     0.00000000
Ru     6.76500000     2.34346474     0.00000000
Ru     9.47100000     2.34346474     0.00000000
Ru     2.70600000     4.68692949     0.00000000
Ru     5.41200000     4.68692949     0.00000000
Ru     8.11800000     4.68692949     0.00000000
Ru    10.82400000     4.68692949     0.00000000
Ru     4.05900000     7.03039423     0.00000000
Ru     6.76500000     7.03039423     0.00000000
Ru     9.47100000     7.03039423     0.00000000
Ru    12.17700000     7.03039423     0.00000000
Ru    -0.00136840     1.56174688     2.06408388
Ru     2.70110963     1.55948900     2.07082718
Ru     5.41086502     1.55837587     2.07711075
Ru     8.11833772     1.55987661     2.07304974
Ru     1.35330500     3.89793198     2.06507491
Ru     4.05502441     3.90676008     2.07711023
Ru     6.76166916     3.90385269     2.07655013
Ru     9.46726783     3.90096451     2.06237094
Ru     2.71405067     6.25017879     2.05595831
Ru     5.41005972     6.25074934     2.07304989
Ru     8.11196643     6.24841311     2.06237112
Ru    10.82004159     6.24695486     2.06200500
Ru     4.05763500     8.59191517     2.06545143
Ru     6.76382622     8.59180107     2.06408414
Ru     9.46435888     8.59689090     2.06507525
Ru    12.18183940     8.59920503     2.05595894
Ru    -0.00041671    -0.00023778     4.12691244
Ru     2.70262597    -0.00553927     4.13281504
Ru     5.41234533     0.01078799     4.14455503
Ru     8.11678033    -0.01272629     4.13662349
Ru     1.34651103     2.34332006     4.13281461
Ru     4.05716591     2.34240814     4.13588907
Ru     6.76097012     2.34098676     4.13848355
Ru     9.47539033     2.34916756     4.13242328
Ru     2.71551362     4.68183464     4.14455349
Ru     5.40783685     4.68467985     4.13848406
Ru     8.09399668     4.67307167     4.13319202
Ru    10.82549671     4.58549753     4.16861608
Ru     4.04736516     7.03570406     4.13662353
Ru     6.77212993     7.03134833     4.13242357
Ru     9.38389895     7.08241392     4.16861369
Ru    12.26422922     7.08075548     4.15948435
Ru    10.99397706     6.34739536     6.04445643
N     11.41671246     6.59145830     7.95662211
N     11.70077919     6.75546121     9.03732173
H      9.52178918     7.23523233     6.02659992
H     11.02668182     4.62852074     6.02660530
\end{verbatim}

\subsubsection*{Product, RuH(NNH)*}
\begin{verbatim}
Lattice vectors (Angstrom):
  a1 = 10.82400017   0.00000000   0.00000000
  a2 =  5.41200009   9.37385475   0.00000000
  a3 =  0.00000000   0.00000000  19.98312841

Ru     0.00000000     0.00000000     0.00000000
Ru     2.70600000     0.00000000     0.00000000
Ru     5.41200000     0.00000000     0.00000000
Ru     8.11800000     0.00000000     0.00000000
Ru     1.35300000     2.34346474     0.00000000
Ru     4.05900000     2.34346474     0.00000000
Ru     6.76500000     2.34346474     0.00000000
Ru     9.47100000     2.34346474     0.00000000
Ru     2.70600000     4.68692949     0.00000000
Ru     5.41200000     4.68692949     0.00000000
Ru     8.11800000     4.68692949     0.00000000
Ru    10.82400000     4.68692949     0.00000000
Ru     4.05900000     7.03039423     0.00000000
Ru     6.76500000     7.03039423     0.00000000
Ru     9.47100000     7.03039423     0.00000000
Ru    12.17700000     7.03039423     0.00000000
Ru    -0.00329431     1.56064088     2.07034459
Ru     2.70606775     1.56309023     2.06870493
Ru     5.41305088     1.56660733     2.07872500
Ru     8.11590394     1.55996611     2.06949850
Ru     1.36678650     3.89668025     2.04363521
Ru     4.06274824     3.90562886     2.07920552
Ru     6.76632811     3.90730742     2.06062807
Ru     9.47116333     3.89318106     2.07061108
Ru     2.69853551     6.25065860     2.07618345
Ru     5.40812238     6.24973077     2.07033470
Ru     8.10627104     6.25707511     2.06979500
Ru    10.81423850     6.24452692     2.06393209
Ru     4.05348689     8.59013082     2.07007626
Ru     6.76082566     8.59186531     2.06983361
Ru     9.46936897     8.61113010     2.04276342
Ru    12.17400999     8.58674072     2.07615966
Ru    -0.00142294     0.00035783     4.13208869
Ru     2.70357919    -0.00388318     4.13305690
Ru     5.40999179     0.02065603     4.13573311
Ru     8.10470339    -0.01550631     4.13813038
Ru     1.34718015     2.34483773     4.13348804
Ru     4.05765539     2.34405777     4.14066056
Ru     6.76878301     2.34646359     4.13618772
Ru     9.47257005     2.31742261     4.13953770
Ru     2.72183688     4.67784387     4.13635245
Ru     5.41506652     4.69036278     4.13658522
Ru     8.12584215     4.69156736     4.20296845
Ru    10.87062050     4.56823466     4.14573834
Ru     4.03662201     7.02871080     4.13853788
Ru     6.74195207     7.04830149     4.13975981
Ru     9.39010702     7.13164387     4.14432568
Ru    12.16076403     7.02265418     4.16465972
Ru    10.07421632     5.82630280     6.04927292
N     10.98951667     6.38257085     7.48790931
N     11.81320032     6.87548677     8.24201327
H      8.49487040     4.89971691     5.95821451
H     11.57305122     6.76076086     9.24758224
\end{verbatim}

\subsection{Extracted fragment (18 atoms)}
\label{si:fragcoords}

The 18-atom extracted fragment, as described in main text Section~\ref{sec:fragextraction} (``Fragment Extraction'').

\subsubsection*{Reactant fragment, RuH$_2$(N$_2$)*}
\begin{verbatim}
Ru     6.76166916     3.90385269     2.07655013
Ru     9.46726783     3.90096451     2.06237094
Ru     8.11196643     6.24841311     2.06237112
Ru    10.82004159     6.24695486     2.06200500
Ru     9.46435888     8.59689090     2.06507525
Ru     6.76097012     2.34098676     4.13848355
Ru     9.47539033     2.34916756     4.13242328
Ru     5.40783685     4.68467985     4.13848406
Ru     8.09399668     4.67307167     4.13319202
Ru    10.82549671     4.58549753     4.16861608
Ru     6.77212993     7.03134833     4.13242357
Ru     9.38389895     7.08241392     4.16861369
Ru    12.26422922     7.08075548     4.15948435
Ru    10.99397706     6.34739536     6.04445643
N     11.41671246     6.59145830     7.95662211
N     11.70077919     6.75546121     9.03732173
H      9.52178918     7.23523233     6.02659992
H     11.02668182     4.62852074     6.02660530
\end{verbatim}

\subsubsection*{Product fragment, RuH(NNH)*}
\begin{verbatim}
Ru     6.76632811     3.90730742     2.06062807
Ru     9.47116333     3.89318106     2.07061108
Ru     8.10627104     6.25707511     2.06979500
Ru    10.81423850     6.24452692     2.06393209
Ru     9.46936897     8.61113010     2.04276342
Ru     6.76878301     2.34646359     4.13618772
Ru     9.47257005     2.31742261     4.13953770
Ru     5.41506652     4.69036278     4.13658522
Ru     8.12584215     4.69156736     4.20296845
Ru    10.87062050     4.56823466     4.14573834
Ru     6.74195207     7.04830149     4.13975981
Ru     9.39010702     7.13164387     4.14432568
Ru    12.16076403     7.02265418     4.16465972
Ru    10.07421632     5.82630280     6.04927292
N     10.98951667     6.38257085     7.48790931
N     11.81320032     6.87548677     8.24201327
H      8.49487040     4.89971691     5.95821451
H     11.57305122     6.76076086     9.24758224
\end{verbatim}

\section{Computational Environment}
\label{si:env}

Software versions for all PySCF-based calculations reported in the main text were verified directly from the program-header output generated on the compute cluster. These calculations used Python~3.14.6 from the Anaconda distribution, PySCF~2.14.0, NumPy~2.5.2, SciPy~1.18.0, and h5py~3.16.0. Additional software used throughout the workflow includes Quantum ESPRESSO\cite{giannozzi2009quantum,giannozzi2017advanced} for periodic DFT calculations, OpenFermion\cite{mcclean2020openfermion} for fermion-to-qubit mapping, PennyLane\cite{bergholm2018pennylane} for ADAPT-VQE optimization, and Forte\cite{evangelista2024forte} for DSRG-MRPT2 calculations.

\section{Supplementary Figures and Tables}
\label{si:figures}

\begin{figure}[htbp]
\centering
\includegraphics[width=0.90\textwidth]{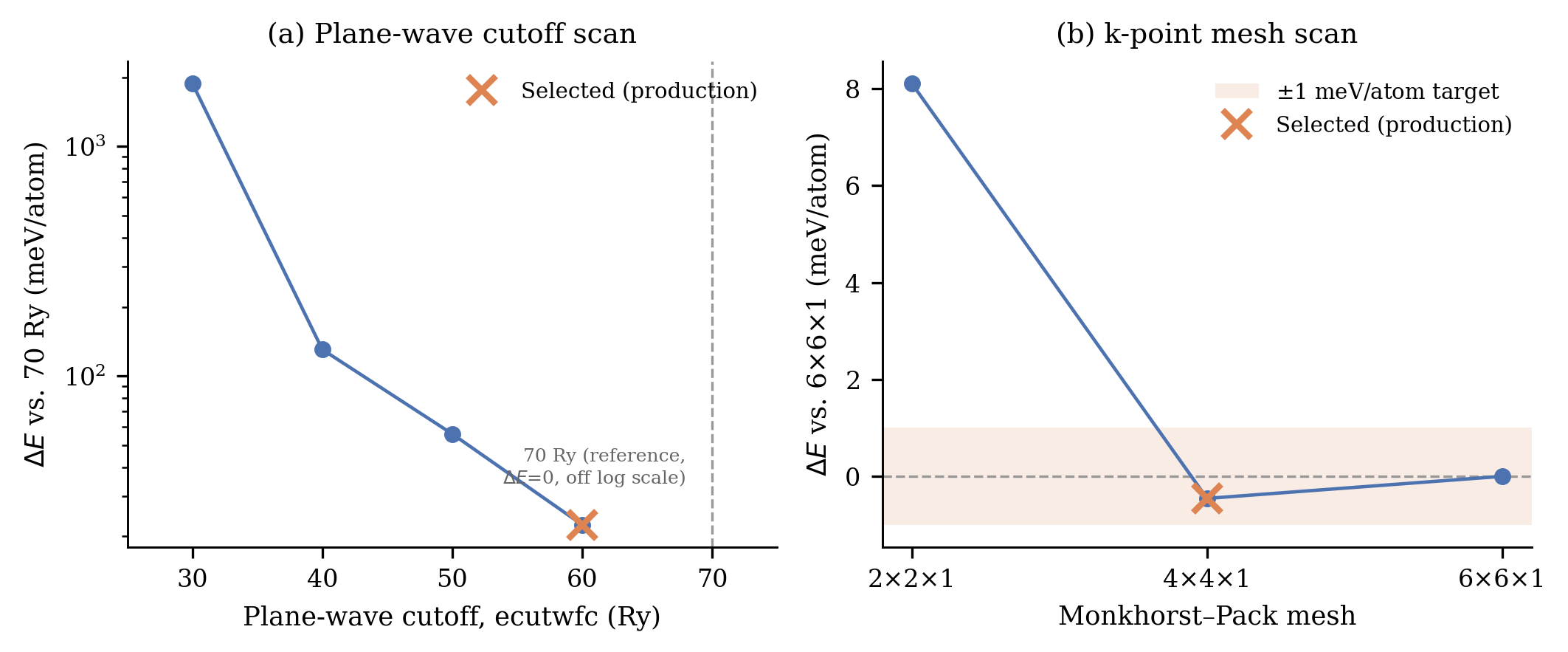}
\caption{DFT convergence on the clean Ru(0001) slab. (a) Plane-wave cutoff scan, energy relative to the 70~Ry reference (dual 12, log scale). (b) Monkhorst-Pack $k$-point mesh scan at the production cutoff (60/480~Ry), energy relative to the 6$\times$6$\times$1 reference; shaded band marks the $\pm$1~meV/atom convergence target. Crosses mark the values adopted for production (60~Ry; 4$\times$4$\times$1); the 70~Ry reference itself corresponds to $\Delta E = 0$ and cannot be shown on the log scale in (a), so its cutoff value is marked with a dashed vertical line instead.}
\label{fig:fig1}
\end{figure}

\begin{figure}[htbp]
\centering
\includegraphics[width=0.55\textwidth]{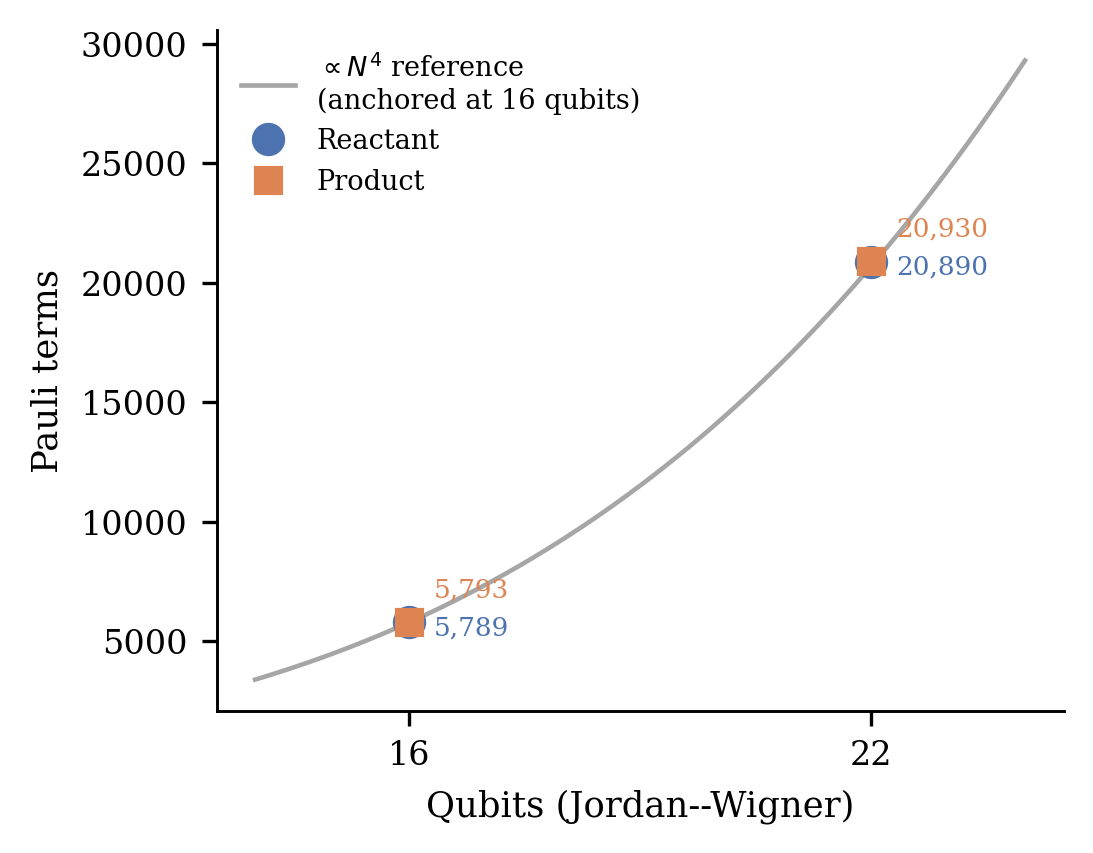}
\caption{Jordan-Wigner Pauli-term count vs.\ qubit count for the two active spaces actually mapped in this work (16 and 22 qubits), against a $\propto N^4$ reference curve ($N$ = spin-orbital count) anchored at the 16-qubit point. Both states' observed term-count ratios (3.61, 3.61) closely track the predicted $(22/16)^4 = 3.57$.}
\label{fig:fig6}
\end{figure}

\FloatBarrier

\subsection{Hardware-Targeted Circuit Resource Estimates}
\label{sec:hwcompare}

The 16-qubit ADAPT-VQE ans\"atz of Section~\ref{sec:methods_adaptvqe} (96 operators for the reactant and 135 for the product) is initially represented as a logical circuit, prior to compilation into the native gate set and connectivity constraints of a specific hardware platform. To estimate the corresponding hardware-level circuit resources, the converged ans\"atze for both reaction states were transpiled (optimization\_level=3) to representative contemporary IBM Heron~r2 and IonQ trapped-ion targets, each compiled through that vendor's own official Qiskit transpilation target rather than a generic or approximated gate set: IBM's Heron~r2 (156~qubits, tunable couplers, native cz two-qubit gate; FakeFez calibration snapshot) and IonQ's native gate set (GPI/GPI2/MS, all-to-all connectivity).\cite{quantum2025scaling,chen2024benchmarking,ionqnativegates}

\begin{figure}[htbp]
\centering
\includegraphics[width=0.75\textwidth]{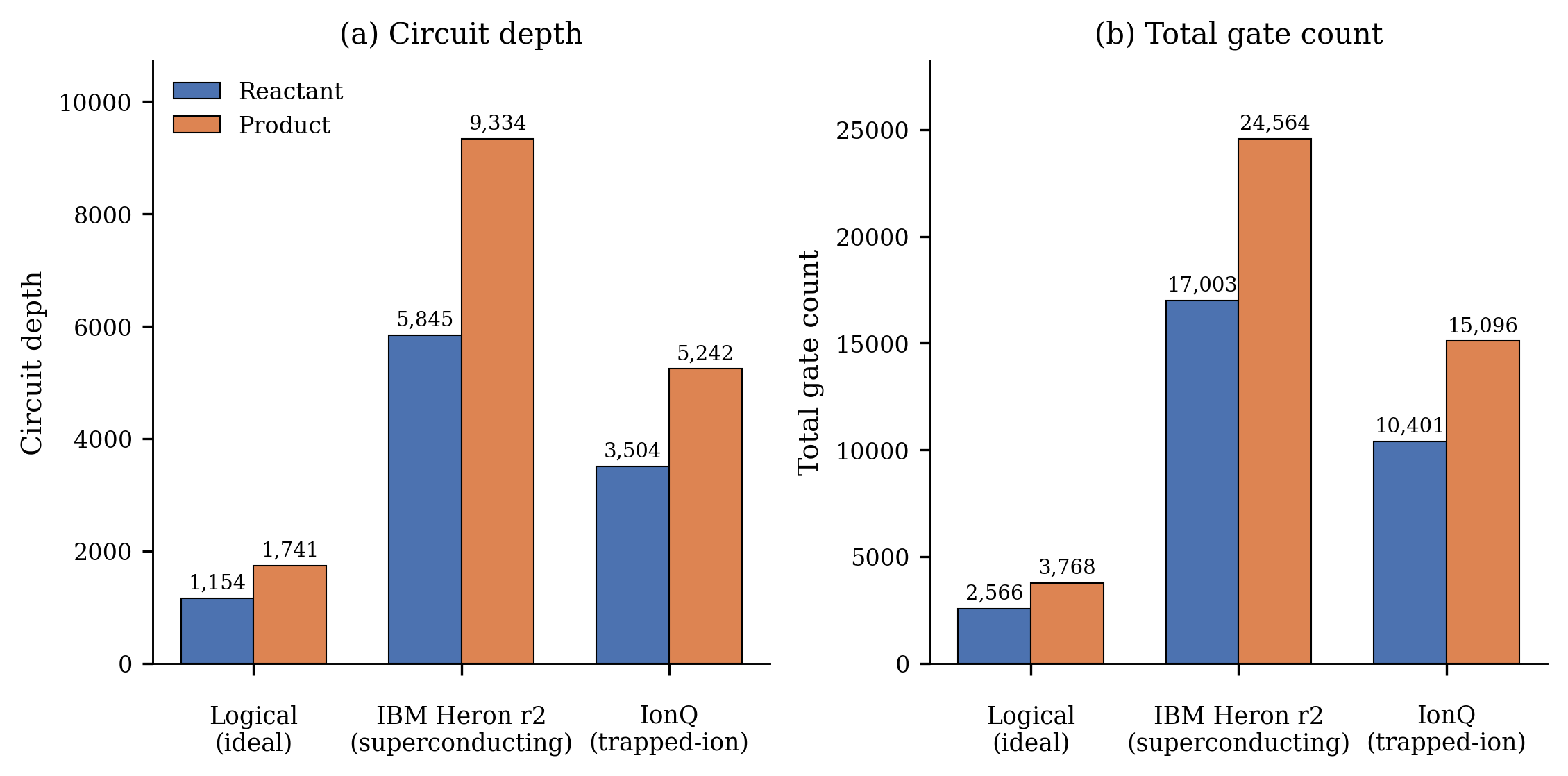}
\caption{Circuit resource requirements for the converged (10e,8o)/16-qubit ADAPT-VQE ans\"atz for the reactant and product, shown at the logical level and after transpilation to representative IBM Heron~r2 and IonQ trapped-ion hardware targets. (a) Circuit depth. (b) Total gate count. The hardware-targeted values are compilation-level resource estimates and reflect the native gate sets, connectivity constraints, and transpilation procedures of the respective targets; they do not represent physical-device execution or a direct comparison of hardware fidelity or performance.}
\label{fig:hwcompare}
\end{figure}

Fig.~\ref{fig:hwcompare} shows the resulting circuit-resource estimates. Circuit depth and total gate count increase substantially relative to the logical circuits for both hardware targets. The compiled resource requirements reflect the combined effects of native-gate decomposition, hardware connectivity, and target-specific transpilation. The IonQ-targeted circuits have lower compiled depth and gate count than the corresponding IBM-targeted circuits for both reaction states. Because the two platforms use different native gate sets, connectivity models, and compilation rules, these differences should not be interpreted as a direct comparison of hardware performance. No noisy simulation or physical-device execution was performed.

\FloatBarrier

\begin{table}[htbp]
\centering
\caption{Reproducibility summary: parameters needed to reproduce the production pipeline, relaxed geometry.}
\begin{tabular}{p{0.40\textwidth}p{0.52\textwidth}}
\toprule
Parameter & Value \\
\midrule
Slab & Ru(0001), $4\times4\times3$ (48 atoms), 11~\AA\ vacuum \\
Plane-wave cutoffs & {ecutwfc} 60~Ry, {ecutrho} 480~Ry (dual 8) \\
$k$-point mesh & $4\times4\times1$ Monkhorst-Pack \\
Smearing & Methfessel-Paxton, $\sigma=0.02$~Ry \\
Dispersion correction & Grimme-D3 \\
Fragment shells & 1 coordination shell (18 atoms) \\
Fragment-DFT validation cell & $30$~\AA\ cubic supercell, $\Gamma$ point \\
Fragment-DFT settings & PBE-D3; same ultrasoft pseudopotentials, 60/480~Ry cutoffs, $\sigma=0.02$~Ry smearing; single-point on extracted geometries \\
Ru basis / ECP & def2-SVP (matching ECP) \\
N, H basis & cc-pVDZ (all-electron) \\
AVAS projector threshold & 0.905 \\
Full active space & (16e,11o), 22 spin-orbitals/qubits (JW) \\
Natural-orbital truncation target & $n_{\mathrm{active}}=8$; verification tolerance 1~kcal/mol \\
Reduced active space & (10e,8o), 16 spin-orbitals/qubits (JW) \\
ADAPT-VQE operator pool & 315 candidates (285 doubles, 30 singles) \\
ADAPT-VQE gradient-norm threshold & $1\times10^{-3}$~Ha \\
ADAPT-VQE optimizer & BFGS \\
ADAPT-VQE max iterations & 100 (reactant); 200 (product) \\
NEVPT2 variant & strongly contracted; full-core and regional-core treatments (Mulliken threshold 0.03 for regional core) \\
DSRG-MRPT2 retained virtuals ($n_{\mathrm{virt}}$) & 50 (not virtual-space converged, Sec.~\ref{sec:dsrgmethod}) \\
DSRG-MRPT2 flow parameter $s$ & $0.3$, $0.5$, $1.0$~Ha$^{-2}$ sensitivity test; default $0.5$ \\

\bottomrule
\end{tabular}
\label{tab:repro}
\end{table}

\FloatBarrier

\begin{table}[htbp]
\centering
\caption{UHF spin-multiplicity scan, relaxed-geometry shell-1 fragment. $\langle S^2 \rangle = 0$ is the ideal value for a true $2S=0$ singlet.}
\begin{tabular}{lrrrl}
\toprule
State & Nominal $2S$ & $E$ (Ha) & $\langle S^2 \rangle$  & SCF converged \\
\midrule
Reactant & 0 & $-1427.10971723$ & $21.597$  &  No \\
Reactant & 2 & $-1426.57815708$ & $15.030$  &  No \\
Reactant & 4 & $-1427.00477771$ & $22.792$  &  No \\
Product  & 0 & $-1427.03670048$ & $20.200$  &  Yes \\
Product  & 2 & $-1426.91727880$ & $19.383$  &  No \\
Product  & 4 & $-1426.98544187$ & $23.240$  &  No \\
\bottomrule
\end{tabular}
\label{tab:spinscan}
\end{table}

\FloatBarrier

\begin{table}[htbp]
\centering
\caption{Plane-wave cutoff scan, bare Ru(0001) slab, fixed dual 12, relative to the 70~Ry reference (48 Ru atoms).}
\begin{tabular}{cc}
\toprule
ecutwfc (Ry) & $\Delta E$ vs.\ 70~Ry (meV/atom) \\
\midrule
30 & $+1881.86$ \\
40 & $+130.33$ \\
50 & $+55.88$ \\
60 (production) & $+22.46$ \\
70 (reference) & $0.00$ \\
\bottomrule
\end{tabular}
\label{tab:cutoffscan}
\end{table}

\FloatBarrier

\begin{table}[htbp]
\centering
\caption{Monkhorst-Pack $k$-point mesh scan, production cutoff (60/480~Ry, dual 8), relative to the 6$\times$6$\times$1 reference.}
\begin{tabular}{cc}
\toprule
Mesh & $\Delta E$ vs.\ 6$\times$6$\times$1 (meV/atom) \\
\midrule
2$\times$2$\times$1 & $+8.108$ \\
4$\times$4$\times$1 (production) & $-0.456$ \\
6$\times$6$\times$1 (reference) & $0.000$ \\
\bottomrule
\end{tabular}
\label{tab:kpointscan}
\end{table}

\FloatBarrier

\begin{table}[htbp]
\centering
\caption{Natural-orbital active-space truncation scan on the converged-geometry (16e,11o) AVAS active space. Error is $|E_{\mathrm{reduced}}^{\mathrm{FCI}} - E_{(16e,11o)}^{\mathrm{CASCI}}|$ for each state, relative to the untruncated (16e,11o) CASCI reference of Section~\ref{sec:notruncresults}.}
\begin{tabular}{ccc}
\hline
$n_{\mathrm{active}}$ & Reactant error (kcal/mol) & Product error (kcal/mol) \\
\hline
4 & $+29.2$ & $+21.9$ \\
5 & $+28.8$ & $+19.8$ \\
6 & $+8.1$ & $+3.8$ \\
7 & $+2.5$ & $+1.7$ \\
8 (adopted) & $+0.002$ & $+0.205$ \\
9 & $+0.001$ & $+0.003$ \\
\hline
\end{tabular}
\label{tab:notrunc}
\end{table}

\FloatBarrier

\begin{longtable}{cccc}
\caption{Eigenvalues of the converged (16e,11o) CASCI one-particle reduced density matrix (Fig.~\ref{fig:fig5}, Section~\ref{sec:notruncresults}), underlying the $n_{\mathrm{active}}=8$ natural-orbital truncation. Orbitals are indexed in descending occupation order; the three orbitals marked ``frozen'' are the ones dropped at $n_{\mathrm{active}}=8$.}
\label{tab:naturalorb}\\
\toprule
Orbital & Reactant occupation & Product occupation & Status ($n_{\mathrm{active}}=8$) \\
\midrule
\endhead
1 & 1.99999980 & 1.99999988 & frozen \\
2 & 1.99999927 & 1.99999729 & frozen \\
3 & 1.99999859 & 1.99981175 & frozen \\
4 & 1.99716932 & 1.99837368 & kept \\
5 & 1.99122941 & 1.99601450 & kept \\
6 & 1.98446374 & 1.98598179 & kept \\
7 & 1.94088772 & 1.98022326 & kept \\
8 & 1.93865573 & 1.93338111 & kept \\
9 & 0.06667279 & 0.06987119 & kept \\
10 & 0.06404877 & 0.01922924 & kept \\
11 & 0.01687485 & 0.01711632 & kept \\
\bottomrule
\end{longtable}

\clearpage

\begin{table}[htbp]
\centering
\caption{NEVPT2 perturber contributions for the plain-AVAS active space using the regional-core treatment at a Mulliken threshold of 0.03.}
\begin{tabular}{ccc}
\toprule
Perturber class & Reactant (Ha) & Product (Ha) \\
\midrule
$S_r$ $(-1)'$ & $-0.06685447$ & $-0.06649540$ \\
$S_i$ $(+1)'$ & $-0.00544961$ & $-0.00620535$ \\
$S_{ijrs}$ $(0)$ & $-1.02671988$ & $-2.11340574$ \\
$S_{ijr}$ $(+1)$ & $-0.00983141$ & $-0.01341704$ \\
$S_{rsi}$ $(-1)$ & $-0.26730842$ & $-0.36421115$ \\
$S_{rs}$ $(-2)$ & $-0.09818903$ & $-0.11118105$ \\
$S_{ij}$ $(+2)$ & $-0.00065266$ & $-0.00066806$ \\
$S_{ir}$ $(0)'$ & $-0.01398143$ & $-0.02905361$ \\
\midrule
Total $E^{(2)}$ & $-1.48898692$ & $-2.70463739$ \\
\bottomrule
\end{tabular}
\label{tab:nevpt2terms}
\end{table}

\FloatBarrier

In Table~\ref{tab:nevpt2terms}, the notation follows the standard NEVPT2 perturber-class labels (Angeli et al.\cite{angeli2001introduction}); the label in parentheses is the change in electron number relative to the CAS reference. The $S_{ijrs}$ (core$\rightarrow$virtual, no active-space change) term dominates and differs by more than a factor of two between the two states, consistent with the anomalously large correction reported in the main text; further analysis of NEVPT2's own internally contracted denominator structure would help clarify its role.

\begin{table}[htbp]
\centering
\caption{DSRG-MRPT2 flow-parameter sensitivity test: reaction energy at three values of $s$, $n_{\mathrm{virt}}=50$, relaxed geometry. With only three points and $n_{\mathrm{virt}}$ itself unconverged, this establishes sensitivity to $s$ rather than a converged scan.}
\begin{tabular}{lrr}
\toprule
$s$ (Ha$^{-2}$) & $\Delta E_{(16e,11o)}$ (kcal/mol) & $\Delta E_{(10e,8o)}$ (kcal/mol) \\
\midrule
0.3 & $+92.6$ & $+83.8$ \\
0.5 & $+70.5$ & $+59.4$ \\
1.0 & $+36.3$ & $+21.1$ \\
\bottomrule
\end{tabular}
\label{tab:dsrgsens}
\end{table}

In Table~\ref{tab:dsrgsens}, the DSRG-MRPT2 row is computed on the relaxed geometry via Forte, using the same regional-core dressed-Hamiltonian construction as the NEVPT2 correction with 50 retained low-lying virtuals ($n_{\mathrm{virt}}=50$). In every case the DSRG-MRPT2 reference energy reproduces the corresponding CASCI energy to $<10^{-8}$~Ha, confirming the dressed Hamiltonian is correct.

\end{document}